\magnification=\magstephalf
\baselineskip=16pt
\parskip=8pt
\rightskip=0.5truecm

\def\a{\alpha}
\def\b{\beta}
\def\d{\delta}
\def\e{\epsilon}

\def\g{\gamma}

\def\r{\rho}
\def\s{\sigma}
\def\t{\tau}

\def\o{\omega}
\def\D{\Delta}
\def\L{\Lambda}
\def\G{\Gamma}
\def\O{\Omega}

\def\del #1{\frac{\partial^{#1}}{\partial\l^{#1}}}

\def\Const{Const\,}
\def\const{const\,}

\def\1{1}

\def\E{I\kern-.25em{E}}
\def\N{I\kern-.22em{N}}
\def\M{I\kern-.22em{M}}
\def\R{I\kern-.22em{R}}
\def\Q{I\kern-.22em{Q}}
\def\Z{{Z\kern-.5em{Z}}}

\def\C{C\kern-.75em{C}}
\def\P{I\kern-.25em{P}}

\def\del{\partial}


\def\NN{{\cal N}}
\def\OO{{\cal O}}

\def\RR{{\cal R}}

\def\UU{{\cal U}}

\def\chap #1{\line{\ch #1\hfill}}

\def\one{\hbox{J}\kern-.2em\hbox{I}}
\def\un #1{\underline{#1}}
\def\ov #1{\overline{#1}}
\def\ba{{\backslash}}
\def\sb{{\subset}}
\def\sp{{\supset}}

\def\em{{\emptyset}}


\newcount\foot
\foot=1
\def\note#1{\footnote{${}^{\number\foot}$}{\ftn #1}\advance\foot by 1}
\def\tag #1{\eqno{\hbox{\rm(#1)}}}
\def\frac#1#2{{#1\over #2}}
\def\text#1{\quad{\hbox{#1}}\quad}

\def\proposition #1{\noindent{\thbf Proposition #1: }}
\def\datei #1{\headline{\rm \the\day.\the\month.\the\year{}\hfill{#1.tex}}}

\def\theo #1{\noindent{\thbf Theorem #1: }}
\def\lemma #1{\noindent{\thbf Lemma #1: }}

\def\proof{{\noindent\pr Proof: }}
\def\proofof #1{{\noindent\pr Proof of #1: }}
\def\endproof{$\diamondsuit$}
\def\remark{{\bf Remark: }}

\def\endproof{$\diamondsuit$}

\font\pr=cmbxsl10 scaled\magstephalf
\font\thbf=cmbxsl10 scaled\magstephalf

\long\def\fussnote#1#2{{\baselineskip=10pt
\setbox\strutbox=\hbox{\vrule height 7pt depth 2pt width 0pt}
\sevenrm
\footnote{#1}{#2}
}}


\font\ch=cmbx12

\font\ftn=cmr8

\font\it=cmti10
\font\bf=cmbx10
\font\srm=cmr5


\overfullrule=0pt
\font\tit=cmbx12
\font\aut=cmbx12
\font\aff=cmsl12
{$  $}
\vskip2truecm
\centerline{\tit  THE CONTINUOUS SPIN RANDOM FIELD MODEL:}
\vskip.2truecm
\centerline{\tit 
FERROMAGNETIC ORDERING IN $d\geq 3$\footnote{${}^*$}{\ftn Work
supported by the DFG
Schwerpunkt `Stochastische Systeme hoher Komplexit\"at'
}}
\vskip2truecm
\vskip.5truecm
\centerline{\aut  Christof K\"ulske\footnote{${}^{1}$}{\ftn
e-mail: kuelske@wias-berlin.de}
}


\vskip.1truecm
\centerline{\aff WIAS}
\centerline{\aff Mohrenstrasse 39}
\centerline{\aff D-10117 Berlin, Germany}
\vskip1.5truecm\rm

\noindent {\bf Abstract:}
We investigate the Gibbs-measures
of ferromagnetically coupled continuous 
spins in double-well potentials subjected
to a random field (our specific example
being the $\phi^4$ theory),
showing ferromagnetic ordering
in $d\geq 3$ dimensions for weak disorder
and large energy barriers. 
We map the random continuous spin distributions 
to distributions for an Ising-spin system 
by means of a single-site
coarse-graining method described by 
local transition kernels. 
We derive a contour-representation for them with notably 
{\it positive} contour activities and prove their Gibbsianness. 
This representation is shown to
allow for application of the discrete-spin
renormalization group developed by 
Bricmont/Kupiainen  implying 
the result in $d\geq 3$.


\vfill
     ${}$
\eject



\magnification=\magstephalf
\baselineskip=16pt
\parskip=8pt
\rightskip=0.5truecm

\def\a{\alpha}
\def\b{\beta}
\def\d{\delta}
\def\e{\epsilon}

\def\g{\gamma}

\def\r{\rho}
\def\s{\sigma}
\def\t{\tau}

\def\o{\omega}
\def\D{\Delta}
\def\L{\Lambda}
\def\G{\Gamma}
\def\O{\Omega}

\def\del #1{\frac{\partial^{#1}}{\partial\l^{#1}}}

\def\Const{Const\,}
\def\const{const\,}

\def\1{1}

\def\E{I\kern-.25em{E}}
\def\N{I\kern-.22em{N}}
\def\M{I\kern-.22em{M}}
\def\R{I\kern-.22em{R}}
\def\Z{{Z\kern-.5em{Z}}}

\def\C{C\kern-.75em{C}}
\def\P{I\kern-.25em{P}}

\def\del{\partial}


\def\NN{{\cal N}}
\def\OO{{\cal O}}

\def\RR{{\cal R}}

\def\UU{{\cal U}}

\def\chap #1{\line{\ch #1\hfill}}

\def\one{\hbox{J}\kern-.2em\hbox{I}}
\def\un #1{\underline{#1}}
\def\ov #1{\overline{#1}}
\def\ba{{\backslash}}
\def\sb{{\subset}}
\def\sp{{\supset}}

\def\em{{\emptyset}}

\def\diam{\hbox{\srm diam}}
\def\Range{\hbox{Range}}


\newcount\foot
\foot=1
\def\note#1{\footnote{${}^{\number\foot}$}{\ftn #1}\advance\foot by 1}
\def\tag #1{\eqno{\hbox{\rm(#1)}}}
\def\frac#1#2{{#1\over #2}}
\def\text#1{\quad{\hbox{#1}}\quad}

\def\proposition #1{\noindent{\thbf Proposition #1: }}
\def\datei #1{\headline{\rm \the\day.\the\month.\the\year{}\hfill{#1.tex}}}

\def\theo #1{\noindent{\thbf Theorem #1: }}
\def\lemma #1{\noindent{\thbf Lemma #1: }}

\def\proof{{\noindent\pr Proof: }}
\def\proofof #1{{\noindent\pr Proof of #1: }}
\def\endproof{$\diamondsuit$}
\def\remark{{\bf Remark: }}

\def\endproof{$\diamondsuit$}

\font\pr=cmbxsl10 scaled\magstephalf
\font\thbf=cmbxsl10 scaled\magstephalf

\long\def\fussnote#1#2{{\baselineskip=10pt
\setbox\strutbox=\hbox{\vrule height 7pt depth 2pt width 0pt}
\sevenrm
\footnote{#1}{#2}
}}


\font\ch=cmbx12

\font\ftn=cmr8

\font\it=cmti10
\font\bf=cmbx10
\font\srm=cmr5



\overfullrule=0pt
\bigskip\bigskip

\chap{I. Introduction}

The study of phase transitions in 
continuous spin lattice models has a long history.  
An important prototypical example of a random model in this class
is the continuous spin random field model,
where ferromagnetically coupled real valued
spins fluctuate in randomly modulated 
local double-well potentials. 

In the present paper we study this model  
for weak disorder in 
dimensions $d\geq 3$ proving ferromagnetic ordering. 
Our aim is more generally to describe an expansion 
method  mapping multiple-well continuous spin 
models to discrete spin models with exponentially
decaying interactions by means of a single-site 
coarse-graining. Then we make use of information 
about the latter ones. This transformation can be regarded as 
an example of a useful (and moreover non-pathological) 
single-site `renormalization group' transformation. 
While it is already interesting in a translation-invariant
situation, it is particularly 
useful for non-translational invariant 
systems since it allows to `factorize' 
the degrees of freedom provided 
by the fluctuations of the spins
around their local minima. 


It is ten years now that the existence of ferromagnetic 
ordering for small disorder at small
temperatures was proved 
for the ferromagnetic random field Ising-model
(with spins $\s_x$ taking values in $\{-1,1\}$)
by Bricmont-Kupiainen [BK1], 
answering a question that had been open 
for long in the theoretical physics community. 
The `converse', namely the a.s. uniqueness 
of the Gibbs-measure in $d=2$ 
was proved later by Aizenman and Wehr [AW].
For an overview about the random field model
from the perspective of theoretical physics, see e.g. [Na].
Given the popularity of {\it continuous} spin
models it is however certainly desirable to have
a transparent method that is able to treat 
the additional degrees of freedom present 
in such a  model. 

Bricmont and Kupiainen introduced in [BK1] 
the conceptually beautiful
method of the renormalization group [RG]
to the rigorous analysis of the low temperature behavior 
of a disordered system,
that turned out to be very powerful in this situation 
although there is no scale-invariance in the 
problem. The heuristic idea is: 
map the initial spin-system onto a coarse-grained one
that appears to be at lower temperature and smaller disorder. 
Then iterate this transformation. 
This idea has to be implemented in a
suitable representation of contours 
(that are the natural variables at low temperatures.)
(For a pedagogical presentation of such a RG
in application to the proof of stability of solid-on-solid
interfaces in disordered media, see also [BoK], [K].)
An alternative treatment of disordered lattice systems with 
finite local spin-space was sketched by Zahradn\' \i k [Z2], 
however also using some iterated coarse graining.

It is clear that also in the more difficult situation
of continuous spins, spatial renormalization 
will be  needed. However, continuous spins being more `flexible' 
than Ising spins make 
it difficult to cut the analysis 
in local pieces. It is then to be expected that the 
difficulties to control the {\it locality} 
of a suitably defined renormalization group 
transformation acting directly 
on continuous spins in a rigorous way 
would blow up tremendously compared 
with the discrete spin case of [BK1]. 
(The amount of technical work needed in their proof 
is already not small!)
For an example of a rigorous construction 
of a RG-group for a continuous spin-lattice system, 
see [Ba1], [Ba2] for the ordered Heisenberg-Ferromagnet.
(This might give some idea of the complexities of such a method.) 

Indeed, despite the conceptual beauty, 
technical difficulties have kept the number of
rigorous applications of the RG to low-temperature 
disordered lattice spin systems limited. 
Moreover, usually a lot of technical work has to be repeated
when extending such a method to a 
more complex situation, while it would be
desirable to make use of older results in 
a more transparent way.     

We will therefore describe a different and more effective 
way to the continuous spin problem: 
1) Construct a single-site `RG'-transformation 
that maps the continuous model to a discrete
one. Obtain bounds on the first in terms
of the latter one. In our specific $\phi^4$ 
double-well situation this 
transformation is just a suitable stochastic mapping
to the sign-field. 
2) Apply the RG group to the discrete model. 
As we will show, the
discrete (Ising-) model in our case 
has a representation as a contour model whose form is 
invariant under the discrete-spin RG that 
was constructed in [BK1].
So we need not repeat the RG analysis
for this part but can apply their results,
avoiding work that has already been done.  

In the last years there has been an ongoing
discussion about the phenomenon of 
RG pathologies. It was  first observed 
by Griffith, Pearce, Israel (and extended 
in various ways by van Enter, Fernandez, Sokal [EFS])
that even very `innocent' transformations like taking
marginals on a sub-lattice of the original lattice
can map a Gibbs-measure of a
lattice spin system to an image measure that need not 
be a Gibbs-measure for any absolutely summable Hamiltonian. 
(See [EFS] for a clear presentation and 
more information about what pathologies can and can not occur,
see also the references given therein.)
On the other hand, as a reaction to this, 
there has been the `Gibbsian restoration program' 
initiated by the late Dobrushin [Do2] whose aim it is to 
exhibit sets of `bad configurations' of measure zero
(w.r.t. the renormalized measure)
outside of which a `renormalized' Hamiltonian
with nicely decaying interactions can be defined. 
This program has been carried out in [BKL]
for a special case (again using RG based on [BK1]).

Since we will be dealing with contour representations
of finite volume measures that provide uniform
bounds on the initial spin system we do not 
have to worry about non-Gibbsianness vs. Gibbsianness
to get our results.
Nevertheless, to put our work in perspective 
with the mentioned discussion, we will 
in fact construct a uniformly convergent `renormalized Hamiltonian' 
for the measure on the sign-field, for {\it all} configurations.  
In other words, there are no pathologies in our 
single-site coarse graining and the situation is as nice 
and simple as it can be.

Let us introduce our model and 
state our main results. 
We are interested in the analysis of the Gibbs measures 
on the state space $\O=\R^{\Z^d}$
of the continuous spin model given by the 
Hamiltonians in finite volume $\L$
$$
\eqalign{
&E^{\tilde m_{\del \L},\eta_{\L}}_{\L}
\left(m_{\L}\right)\cr
&=\frac{q}{2}\sum_{{\{x,y\}\sb \L}\atop {d(x,y)=1}}\left(
m_x-m_y
\right)^2+
\frac{q}{2}\sum_{{x\in \L; y\in \del \L}
\atop{d(x,y)=1}}\left(
m_x-\tilde m_y
\right)^2 + \sum_{x\in \L}V(m_x)- \sum_{x\in \L}\eta_x m_x
}
\tag{1.1}
$$
for a configuration $m_{\L}\in \O_{\L}= \R^{\L}$
with boundary condition $\tilde m_{\del \L}$.
Here we write $\del \L=\{x\in \L^c;\exists y\in \L:  d(x,y)=1\}$
for the outer boundary of a set $\L$ where 
$d(x,y)=\Vert x-y\Vert_1$ is the $1$-norm on $\R^d$.
$q\geq 0$ will be small. 
Given its history and its popularity we will 
consider mainly the example of the well-known double-well 
$\phi^4$-theory. As we will see during
the course of the proof, there 
is however nothing special about this choice. 
We use the normalization where the  minimizers are $\pm m^*$,
the curvature in the minima is $1$, 
and the value of the potential in the minima is zero and write 
$$
\eqalign{
&V(m_x)
=\frac{\left(m_x^2-(m^*)^2\right)^2}{8 {m^*}^2}
}
\tag{1.2}
$$
where the parameter $m^*\geq 0$ will be large. 
We consider i.i.d. random fields 
$\left(\eta_x\right)_{x\in \Z^d}$ that satisfy

\item{(i)} $\eta_x$ and $-\eta_x$ have the same distribution

\item{(ii)} $\P\left[\eta_x\geq t\right]\leq e^{-\frac{t^2}{2\s^2}}$

\item{(iii)} $|\eta_x|\leq \d$ 

\noindent where $\s^2\geq 0$ is sufficiently small. 
The assumption (iii) of having uniform bounds
is not essential for the problem of stability of the
phases but made to avoid uninteresting
problems with our transformation and keep 
things as transparent as possible.

The finite volume Gibbs-measures $\mu_{\L}^{\tilde m_{\del \L},\eta_{\L}}$
are then defined as usual through the expectations
$$
\eqalign{
&\mu_{\L}^{\tilde m_{\del \L},\eta_{\L}}(f)
=\frac{1}{Z_{\L}^{\tilde m_{\del \L},\eta_{\L}}}
\int_{\R^{\L}}dm_{\L}f\left(m_{\L},\tilde m_{\L^c} \right)
e^{-E^{\tilde m_{\del \L},\eta_{\L}}_{\L}
\left(m_{\L}\right)}
}
\tag{1.3}
$$
for any bounded continuous $f$ on $\O$  with 
the partition function
$$
\eqalign{
&Z_{\L}^{\tilde m_{\del \L},\eta_{\L}}
=\int_{\R^{\L}}dm_{\L}
e^{-E^{\tilde m_{\del \L},\eta_{\L}}_{\L}
\left(m_{\L}\right)}
}
\tag{1.4}
$$
We look in particular at the measures 
with boundary condition $\tilde m_x=+m^*$ (for all $x\in \Z^d$) 
in the positive minimum of the
potential, for which we write 
$\mu_{\L}^{+m^*, \eta_{\L}}$.

To prove the existence of a phase transition 
we will show that, for 
a suitable range of parameters, 
with large probability 
w.r.t. the disorder, the Gibbs-expectation 
of finding the field left to the positive well
is very small. 
Indeed, we have as the main result

\theo{1}{\it 
Let $d\geq 3$ and assume the conditions (i),(ii),(iii)
with $\s^2$ small enough. 
Then, for any (arbitrarily small) $\g>0$, 
there exist $q_0>0$ (small enough), $\d_0,\d_1>0$ (small enough),  
$\t_0$ (large enough)
such that, whenever $\d\leq \d_0$,
$q(m^*)^2\geq \t_0$ and $q (m^*)^{\frac{2}{3}}\leq \d_1$
we have that
$$
\eqalign{
&\P\left[
\limsup_{N\uparrow\infty}
\mu_{\L_N}^{+m^*, \eta_{\L_N}}
\left[m_{x_0}\leq \frac{m^*}{2}\right]\geq  \g\right]
\leq  e^{-\frac{\const}{\s^2}} 
\cr
}
\tag{1.5}
$$
for an increasing sequence of cubes $\L_N$. 
}


\noindent{\thbf Remark: } Note that the quantity $q (m^*)^2$ 
gives the order of magnitude of the minimal 
energetic contribution
of a nearest neighbor pair of spins with opposite
signs to the Hamiltonian (1.1); 
it will play the role of a (low temperature) Peierls constant. 
Smallness of $q$ (to be compared with 
the curvature unity in the minima of the potential) 
is needed to ensure a fast decay of correlations
of the thermal fluctuations around the minimizer 
in a given domain. 
The stronger conditions on the 
smallness, $q \leq \const (m^*)^{-\frac{2}{3}}$,
however is needed in our approach 
to ensure the positivity and smallness
of certain anharmonic corrections.

Let us now define the transition kernel 
$T_x\left(\, \cdot\,\bigl| \,\cdot\,\right)$ from 
$\R$ to $\{-1,1\}$ we use 
and explain why we do it.  
Put, for a continuous spin $m_x\in \R$,
and an Ising spin $\s_x\in \{-1,1\}$
$$
\eqalign{
&T_x\left(
\s_{x}\Bigl |m_x
\right):=\frac{1}{2}\left(1+\s_x\tanh\left(a m^* m_x \right)\right)
}
\tag{1.6}
$$
where $a\geq 1$, close to $1$, will have to be chosen later
to our convenience.  
In other words, 
the probability that a continuous spin 
$m_x$ gets mapped to its sign is given by 
$\frac{1}{2}\left(1+\tanh\left(a m^* \left|m_x\right| \right)\right)$
which converges to one for large $m^*$.
The above kernel 
defines a joint probability distribution 
$\mu_{\L}^{\tilde m_{\L^c},\eta_{\L}}(dm_{\L})
T\left(d\s_{\L}\bigl| m_{\L} \right)
$ on $\R^\L\times \{-1,1\}^\L$
whose non-normalized density is given by 
$$
\eqalign{
&e^{-E^{\tilde m_{\del \L},\eta_{\L}}_{\L}
\left(m_{\L}\right)} \prod_{x\in \L}T_x(\s_x\bigl| m_x)
}
\tag{1.7}
$$
Its marginal on the Ising-spins $\s_{\L}$
$$
\eqalign{
&\left(T\left(\mu_{\L}^{\tilde m_{\del\L},\eta_{\L}}\right)
\right)(d\s_{\L})
:=\int_{\R^{\L}}\mu_{\L}^{\tilde m_{\del\L},\eta_{\L}}(dm_{\L})
T\left(d\s_{\L}\bigl| m_{\L} \right)
}
\tag{1.8}
$$
will be the main object of our study. 

To prove the existence of a phase transition 
stated in Theorem 1 
we will have to deal only with 
finite volume contour representations of (1.8),
as given in Proposition 5.1. Nevertheless,
it is perhaps most 
instructive to present the following infinite volume result
in the Hamiltonian formulation to 
explain the nature of the transformation. 

\theo{2}{\it  Assume the hypothesis
of Theorem 1 and let $\eta$ be any fixed  realization of 
the disorder. 
Suppose that  $\mu^{\eta}$ is a continuous spin Gibbs-measure 
obtained as a weak limit of 
$\mu_{\L}^{\tilde m_{\del \L},\eta_{\L}}$
along  a sequence of cubes $\L$ 
for some boundary condition 
$\tilde m\in \{-m^*,m^*\}^{\Z^d}$. 
Then, for a suitable choice of 
the parameter $a\geq 1$ (close to $1$) in the kernel $T$
the following is true. 

The measure $T\left(\mu^{\eta}\right)$
on $\{-1,1\}^{\Z^d}$ is a Gibbs measure
for the absolutely summable Ising-Hamiltonian
$$
\eqalign{
&H_{\srm Ising}^{\eta}\left(\s\right)\cr
&=-\frac{a^2 (m^*)^2}{2}\sum_{x,y}\left( 
a-q\D_{\Z^d}\right)^{-1}_{x,y}\s_{x}\s_{y}
- a m^* \sum_{x}\left(a-q\D_{\Z^d}\right)^{-1}_{x,y}\eta_x \s_x
-\sum_{C:|C|\geq 2}\Phi_{C}\left(\s_C;\eta_C\right)
}
\tag{1.9}
$$
where $\D_{\Z^d}$ is the lattice 
Laplacian in the infinite volume, i.e.
$\D_{\Z^d;x,y}=1$ iff $x,y\in V$ are nearest neighbors,
$\D_{\Z^d;x,y}=-2d$ iff $x=y$ and $\D_{\Z^d;x,y}=0$ else.

The many-body potentials 
are symmetric under joint flips of spins and random-fields,
$\Phi_C(\s_C,\eta_{C})=\Phi_C(-\s_C,-\eta_{C})$,
and translation-invariant under joint lattice-shifts. 
They obey the uniform bound
$$
\eqalign{
&\left|\Phi_C(\s_{C},\eta_C)\right|\leq e^{-\tilde \g|C|}
\cr
}
\tag{1.10}
$$
with a positive constant $\tilde \g$.
}

\bigskip

\noindent{\thbf Remark 1:} As in Theorem 1, $\tilde \g$
can be made arbitrarily small by choosing $q_0,\d_0,\d_1$
small and $\t_0$ large. 
More information about estimates on the value 
of $\g$ and $\tilde \g$ can in principle be deduced
from the proofs. 

\noindent{\thbf Remark 2:}
By imposing the smallness of $\d$ we 
exclude pathologies due to exceptional
realizations of the disorder variable $\eta$ 
(`Griffiths singularities') in the transformation $T$.
(We stress that this does not simplify 
the physical problem of the study of the low-temperature
phases which is related to the 
study of the formation of large contours.)
Starting from the joint distribution (1.7)
it is natural to consider the distribution 
of continuous spins {\it conditional} on the Ising spins; 
here the Ising spins $\s_x$ 
will play the role of a second sort of external fields.
Then, as it was explained in [BKL],
possible pathologies in the transformation $T$ would be 
analogous to Griffiths-singularities
created by pathological Ising configurations. 
In this sense, Theorem II states that there are neither
Griffiths singularities of the first type (w.r.t. $\eta$)
nor the second type (w.r.t $\s$).
The treatment of unbounded random fields
would necessitate the analysis of so-called `bad regions' 
in space (where the realizations of the random
fields are anamolously large). 
This should be possible but would however obscure the nature 
of the transformation $T$.

Let us now motivate the form of $T_x$ and comment
on the structure of the Hamiltonian.
Introducing quadratic potentials, centered at $\pm m^*$,
$$
\eqalign{
&Q^{\s_x}\left(
m_x\right):= \frac{a }{2}\left(m_x-\s_x m^*\right)^2+b \cr
}
\tag{1.11}
$$
with $b>0$ (close to zero) 
to be chosen later, we can rewrite
the transition kernel in the form 
$$
\eqalign{
&T_x\left(
\s_{x}\bigl |m_x\right)
=\frac{e^{-Q^{\o_x}\left(
m_x\right)}}
{\sum_{\bar\o_x=\pm 1}e^{-Q^{\bar\o_x}\left(
m_x\right)}}
}
\tag{1.12}
$$
The crucial point is that 
the joint density (1.7) contains a product over $x$ over
the quantities
$$
\eqalign{
&e^{-V(m_x)}T_x\left(
\s_{x}\Bigl |m_x\right)
=e^{-Q^{\s_x}\left(
m_x\right)}\left(1+w(m_x)\right) 
}
\tag{1.13}
$$
where, using (1.12), we can write the remainder in the form 
$$
\eqalign{
&\left(1+w(m_x)\right)
:=\frac{e^{-V(m_x)}}
{\sum_{\bar\s_x=\pm 1}e^{-Q^{\bar\s_x}\left(
m_x\right)}} 
}
\tag{1.14}
$$
Now, if the initial potential $V(m_x)$ is sufficiently 
Gaussian around its minima
and the quadratic potential $Q^{\s_x}$
is suitably chosen, $w(m_x)$ should be {\it small } in some sense.
If $w(m_x)$ were even zero, we would be left 
with $\s_{\L}$-dependent Gaussian integrals
that can be readily carried out. They lead
to the first two terms in the Ising-Hamiltonian (1.9),
containing only pair-interactions. 
This can be understood by a formal computation. 
The modification of the measure for `small' $w(m_x)$ 
then gives rise indeed to exponentially decaying 
many-body interactions, as one could naively 
hope for. 

Expanding $\prod_{x\in \L}\left(1+w(m_x)\right)$
then leads in principle to an expansion around
a Gaussian field.\footnote{$^1$}
{\ftn The author is grateful to M. Zahradn\' \i k 
for pointing out the idea to decompose $e^{-V(m_x)}$
into a sum of two Gaussians and a remainder term
that should be expanded. However, contrary to [Z3] 
we write the remainder in a  
multiplicative form which allows for
the transition kernel interpretation.} 
One problem with this direct treatment is however that 
resulting contour activities will in general be nonnegative only
if $w(m_x)\geq 0$ for all $m_x$.
But note that the latter can only be true for the narrow
class of potentials
such that $V(m_x)\leq \Const m_x^2$ for large $|m_x|$.
Thus,  $w(m_x)$ will have to become negative for some 
$m_x$ e.g. for $V$ compact support or in the $\phi^4$-theory. 
While it is not necessary to have positive 
contour activities for some applications (see [BChF],[Z3]) 
it  is crucial for the random model:
A RG, as devised in [BK1], 
needs non-negative contour weights.\footnote{$^2$}{\ftn
Vaguely speaking, the method 
keeps lower bounds on the energies of all configurations, 
but also upper bounds on the energies of some
configurations (that are candidates for the true groundstates).
This can be seen nicely in the 
groundstate-analysis of the models treated in [BoK]. 
To do an analogue of this 
for finite temperatures, non-negative
(probabilistic) contour weights are necessary in this framework.} 
We are able to solve this problem and define positive
effective anharmonic weights
by a suitable resummation and careful 
choice of the parameters $a,b$ of the quadratic
potential $Q^{\s_x}$; these will be kept fixed. 
This choice is the only
point of the proof that has to be adapted
to the specific form of the initial potential $V$. 
Later the positivity of weights will also be used 
for the control of the original measure in terms
of the Ising-measure (see Proposition 5.2).

In Chapter II it is shown 
how non-negative effective anharmonic weights
obeying suitable Peierls bounds
can be defined.
Chapter III finishes the control of the anharmonicity
around the Ising model arising
from the purely Gaussian theory (i.e. $w(m_x)\equiv 0$)
in terms of a uniformly convergent expansion. 
Chapter IV treats the simple but instructive case of 
the Ising field without the presence of anharmonicity, 
showing the emergence 
of (generalized) Peierls bounds on Ising contours.  
In Chapter V we obtain our final contour
model for the full theory and prove Theorem 1 and Theorem 2.
The Appendix collects
some facts about Gaussian random fields
and random walk expansions we employ.

\bigskip

\chap{Acknowledgments:}
The author thanks A.Bovier and M.Zahradn\' \i k
for interesting discussions and suggestions. 
This work was supported by the DFG,
Schwerpunkt `Stochastische Systeme hoher Komplexit\"at'.





\bigskip\bigskip
\overfullrule=0pt

\chap{II. Anharmonic contours with positive weights\hfill}

We will explain in this Chapter how (preliminary) 
`anharmonic contours' with
`anharmonic weights' that are non-negative
and obey a Peierls estimate can be constructed.
We start with a combinatorial Lemma 2.1.
and a suitable organization 
of the order of Gaussian integrations appearing
to derive algebraically the representation of Lemma 2.3. 
We will make no specific assumptions
about the potential at this point 
that should however 
be thought to be symmetric `deep' double-well. 
Our later treatment is valid once
we have the properties of 
`positivity' and `uniform Peierls condition
of anharmonic weights' that are introduced
in (2.19) and (2.20). These are then verified for the
$\phi^4$-theory in an isolated 
part of the proof that can be adapted
to specific cases of interest. 

We will have to deal with the interplay
of three different fields: continuous spins $m_x$
(to be integrated out), Ising spins $\s_x$ 
and (fixed) random fields $\eta_x$,
subjected to various boundary conditions
in various volumes. In some sense, the general
theme of the expansions to come is: 
keep track of the locality  
of the interaction of these fields in the right way. 
For the sake of 
clarity we found it more appropriate in 
this context to keep a notation that
indicates the dependence on these quantities 
in an explicit way
in favor of a more space-saving one.  

Now, since  we are interested here 
in a contour-representation of the image measure 
$T\left(\mu_{\L}^{\tilde m_{\del\L},\eta_{\L}}\right)$
under the stochastic transformation (1.6), let us 
look at the non-normalized weights on Ising-spins
given by  
$$
\eqalign{
&Z_{\L}^{\tilde m_{\del \L},\eta_{\L}}(\s_{\L})
:=\int_{\R^{\L}}dm_{\L}
e^{-E^{\tilde m_{\del \L},\eta_{\L}}_{\L}
\left(m_{\L}\right)}\prod_{x\in \L}T_x(\s_x\bigl| m_x)\cr
}
\tag{2.1}
$$
so that we get the desired Ising-probabilities 
dividing by 
$Z_{\L}^{\tilde m_{\del \L},\eta_{\L}}=\sum_{\s_{\L}\in \{-1,1\}^{\L}}
Z_{\L}^{\tilde m_{\del \L},\eta_{\L}}(\s_{\L})$.

To describe our expansions conveniently 
let us define the following quadratic continuous-spin 
Hamiltonians, that are made to collect 
the quadratic terms that arise from the use 
of (1.13) to the above integral. 
We write, for finite volume $V\sb \Z^d$,
$$
\eqalign{
&H^{\tilde m_{\del V},\eta_{V}, \s_{V}}_{V}
\left(m_{V}\right)\cr
&=\frac{q}{2}\sum_{{\{x,y\}\sb V}\atop {d(x,y)=1}}\left(
m_x-m_y
\right)^2+
\frac{q}{2}\sum_{{x\in V; y\in \del V}
\atop{d(x,y)=1}}\left(
m_x-\tilde m_y
\right)^2 + \frac{a}{2}\sum_{x\in V}\left(m_x-m^*\s_x\right)^2 
-\sum_{x\in V}\eta_x m_x
}
\tag{2.2}
$$
Here and throughout the paper we always write $\del G$
for the outer boundary {\it inside} $\L$, i.e.
$\del G=\{x\in \L\cup B^c; d(x,G)=1\}$.
The notion `nearest neighbor' is always meant in the usual sense 
of the $1$-norm.
The fixed Ising-spin $\s_V\in \{-1,1\}^V$ thus signifies the 
choice of the well at each site.
From the point of view of the continuous fields
it is just another parameter.

With this definition we can write 
the non-normalized Ising-weights (2.1) in the form 
$$
\eqalign{
&Z_{\L}^{\tilde m_{\del \L},\eta_{\L}}(\s_{\L})
=e^{-b|\L|}\int_{\R^{\L}}dm_{\L}e^{
-H^{\tilde m_{\del \L},\eta_{\L},\s_{\L}}_{\L}
\left(m_{\L}\right)}
\prod_{x\in \L}\left(1 +w(m_x)\right)
}
\tag{2.3}
$$
If the $w(m_x)$ were identically zero, we would
be left with purely Gaussian integrals
over Ising-spin dependent quadratic expressions. 
This Gaussian integration can 
be carried out and yields
$$
\eqalign{
&\int_{\R^{\L}}dm_{\L}e^{
-H^{\tilde m_{\del \L},\eta_{\L},\s_{\L}}_{\L}
\left(m_{\L}\right)}
=C_{\L}\times e^{
-\inf_{m_{\L}\in \R^{\L}}H^{\tilde m_{\del \L},\eta_{\L},\s_{\L}}_{\L}
\left(m_{\L}\right)}
}
\tag{2.4}
$$
with a constant $C_{\L}$ that does not depend on 
$\s_{\L}$ (and $\eta_{\L}$). 
The latter fact is clear since 
$\s_{\L}$ (and $\eta_{\L}$) only couple as linear 
terms (`magnetic fields') to $m_{\L}$
while they do not influence the quadratic terms.
Note the pleasant fact 
that no spacial decomposition of the Gaussian
integral is needed here and no complicated boundary terms arise.

Now the minimum of the continuous-spin Hamiltonian
in the expression on the r.h.s. of (2.4) 
provides weights for an effective random field 
Ising model for the spins $\s_{\L}$; 
its (infinite volume) 
Hamiltonian is given by the first two terms in (1.9).
The  treatment of this model 
is much simpler than that of the 
full model; all this will be postponed to Chapter IV.
There it is discussed in detail
how this model can be transformed into 
a disordered contour model by a mixed low- and
high-temperature expansion. However, since this model provides
the main part of the final
contour model that is responsible for the ferromagnetic
phase transition some readers might want to 
take a look to Chapter IV to understand the 
form of our final contour-representation in a simpler situation. 

Our present aim now is however to show how the anharmonic 
perturbation induced by the $w$-terms can be treated
as a positive-weight perturbation of the purely Gaussian
model.

Let $U= U^+\cup (-U^+)\sb \R$,  
where $U^+$ is a suitable `small' neighborhood of 
the positive minimizer of the potential
$m^*$ that will be determined later and that
will depend on the specific form of the potential. 
The first key step to define non-negative activities
is now to use the following
combinatorial identity on the set $\UU=
\{x\in \L; m_x\in U\}$.

\lemma{2.1}{\it Let $\L\sb \Z^d$ be finite and connected. 
For any set $\UU\sb \L$ we can 
write the polynomial $\prod_{x\in \L}\left(1 + w_x\right)$
in the  $|\L|$
variables $\left(w_x\right)_{x\in \L}$
in the form 
$$
\eqalign{
&\prod_{x\in \L}\left(1 + w_x\right)\cr
&=1+\sum_{G:\em \neq G \subset \L } 
\prod_{{G_i}\atop{\hbox{ conn.cp of }G}}
\prod_{x\in \del G_i}1_{x\in \UU}\left[
\prod_{x\in G_i}\left(1_{x\not\in \UU}+w_x \right)
-\prod_{x\in G_i}1_{x\not\in \UU}
\right]
}
\tag{2.5}
$$
}

The proof is given at the end of this chapter.
Application of Lemma 2.1. gives us the expansion
$$
\eqalign{
&e^{b|\L|}Z_{\L}^{\tilde m_{\del \L},\eta_{\L}}(\s_{\L})
=\int_{\R^{\L}}dm_{\L}e^{
-H^{\tilde m_{\del \L},\eta_{\L},\s_{\L}}_{\L}
\left(m_{\L}\right)}\cr
&+\sum_{G:\em \neq G \subset \L } 
\int_{\R^{\L}}dm_{\L}e^{
-H^{\tilde m_{\del \L},\eta_{\L},\s_{\L}}_{\L}
\left(m_{\L}\right)}\cr
&\quad\times \prod_{{G_i}\atop{\hbox{ conn.cp of }G}}
\prod_{x\in \del G_i}1_{m_x\in U}\left[
\prod_{x\in G_i}\left(1_{m_x\not\in U}+w(m_x) \right)
-\prod_{x\in G_i}1_{m_x\not\in U}
\right]
}
\tag{2.6}
$$
Note that the expression under the integral factorizes
over connected components of $\ov {G}:=G\cup \del G$.

To introduce the anharmonic (preliminary) weights we need a little
preparation. 
To avoid unnecessary 
complications in the expansions it is important to organize 
the Gaussian integral in the following conceptually simple 
but useful way:
We decompose the nonnormalized Gaussian expectation over 
the terms in the last line
into an {\it outer} integral over $m_{\del G}$ 
and a `conditional integral' over $m_{\L\ba\del G}$ 
{\it given} $m_{\del G}$. The latter integral
factorizes of course over connected components
of $\L\ba\del G$; in particular the integrals
over $\L\ba\ov G$ and $G$ become conditionally independent.
W.r.t. this decomposition they appear in a symmetric way. 

To write down the explicit formulae we need for that
we introduce

{\noindent{\thbf Some notation: }}{
The   $V\times V$-matrix 
$\D_{V}$ is the lattice Laplacian with Dirichlet boundary 
conditions on $V\sb \L$, i.e.
$\D_{V;x,y}=1$ iff $x,y\in V$ are nearest neighbors,
$\D_{V;x,y}=-2d$ iff $x=y\in V$ and $\D_{V;x,y}=0$ else.
$\Pi_{V}$ is the projection operator onto $\O_{V}$
(in short: onto $V$), i.e. $\Pi_{V; x,y}=1_{x=y\in V}$. 
We also use the redundant but intuitive
notations $m_{\L}|_{V}\equiv\Pi_{V}m_{\L}\equiv m_{V}$ 
for the same thing.
$1_{V}$ is the vector in $\R^{\L}$ given by $1_{V;x}=1_{x\in V}$.
For disjoint 
$V_1,V_2\sb \L$ we write $\del_{V_1,V_2}$ 
for the matrix with entries $\del_{V_1,V_2;x,y}=1$ iff $x\in V_1$,
$y\in V_2$ are nearest neighbors and $\del_{V_1,V_2;x,y}=0$ else.
We write
$R_{V}:=\left (c- \D_{V}\right)^{-1}$
for the corresponding resolvent in the volume $V$.
Here and later we put $c=\frac{a}{q}$.}

For the sake of clarity we keep (at least for now)
the dependence of all quantities on 
continuous spin-boundary conditions,
random fields,  Ising-spins, as superscripts.
Then we have

\lemma{2.2}{\it For any subset $G\sb \L$ the 
random quadratic Hamiltonians (2.2) have the decomposition 
$$
\eqalign{
&H^{\tilde m_{\del \L},\eta_{\L}, \s_{\L}}_{\L}
\left(m_{\L}\right)\cr
&=\D H^{\tilde m_{\del \L},
\eta_{\L},\s_{\L}}_{\del G,\L}\left(m_{\del G} \right)
+\D H^{\tilde m_{\del \L},
m_{\del G},\eta_{\L\ba\del G},\s_{\L\ba\del G}}
_{\L\ba\del{G}}\left(m_{\L\ba\del{G}} \right)
+\inf_{m'_{\L}} H^{\tilde m_{\del \L},\eta_{\L}, 
\s_\L}_{\L}\left(m'_{\L}\right)\cr
}
\tag{2.7}
$$
Here the `fluctuation-Hamiltonians' are given by 
$$
\eqalign{
&\D H^{\tilde m_{\del \L},
\eta_{\L},\s_{\L}}_{\del G,\L}
\left(m_{\del G} \right)\cr
&=\frac{1}{2}
<\left(m_{\del G}
-m_{\L}^{\tilde m_{\del \L},
\eta_{\L},\s_{\L}}\bigl|_{\del G} \right),
\left(\Pi_{\del G}\left(a-q\D_{\L}\right)^{-1}\Pi_{\del G}\right)^{-1}
\left(m_{\del G}
-m_{\L}^{\tilde m_{\del \L},
\eta_{\L},\s_{\L}}\bigl|_{\del G} \right)>_{\del G}\cr
}
\tag{2.8}
$$
and the `conditional fluctuation-Hamiltonian' 
(i.e. conditional on $m_{\del G}$)
$$
\eqalign{
&\D H^{\tilde m_{\del \L},m_{\del G},\eta_{\L\ba\del G},\s_{\L\ba\del G}}
_{\L\ba\del{G}}\left(m_{\L\ba\del{G}} \right)\cr
&=\frac{1}{2}<\left(m_{\L\ba\del{G}}-
m_{\L\ba\del G}^{\tilde m_{\del \L},m_{\del G},
\eta_{\L\ba\del G},
\s_{\L\ba\del G}}\right),
\left(a-q\D_{\L\ba\del{G}}\right)\left(m_{\L\ba\del{G}}-
m_{\L\ba\del G}^{\tilde m_{\del \L},m_{\del G},
\eta_{\L\ba\del G},
\s_{\L\ba\del G}}\right)>_{\L\ba G}
}
\tag{2.9}
$$
As centerings are occuring: the `global minimizer' 
$$
\eqalign{
&m_{\L}^{\tilde m_{\del \L},
\eta_{\L},
\s_{\L}}
=R_{\L}\left(
c m^*\s_{\L}
+\frac{\eta_{\L}}{q}
+\del_{\L,\del \L}\tilde m_{\del \L}
\right)\cr
}
\tag{2.10}
$$
and the `conditional minimizer'
$$
\eqalign{
&m_{\L\ba\del G}^{\tilde m_{\del \L},m_{\del G},
\eta_{\L\ba\del G},
\s_{\L\ba\del G}}\cr
&=R_{\L\ba\del G}\left(
c m^*\s_{\L\ba\del G}
+\frac{\eta_{\L\ba\del G}}{q}
+\del_{\L\ba\del G,\del G}m_{\del G}
+\del_{\L\ba\del G,\del \L}\tilde m_{\del \L}
\right)\cr
}
\tag{2.11}
$$

}

The proof is a consequence of Appendix Lemma A.1(iii)
which is just a statement about symmetric 
positive definite matrices.
Lemma 2.2 can be seen as an explicit expression of the 
compatibility property for the Gaussian 
local specifications defined thru the Hamiltonian
(2.7) in the volumes $\L\ba\del {G}\sb \L$.
Indeed, the Gaussian measure defined
with the quadratic form (2.8) describes 
the distribution on $\L$ projected onto $\del G$.
(Since we will use this formula
later for subsets of $\L$ it is convenient to 
make the $\L$ explicit at this point, too.)
The Gaussian measure on $\L\ba\del G$ 
defined with (2.9) is the conditional measure given 
$m_{\del G}$.

We like to stress 
the following decoupling properties of the conditional expressions. 
Equation (2.11) for the conditional minimizer decouples 
over connected components $V_i$ of $\L\ba\del G$ since
the resolvent $R_{\L\ba\del G}$ is just 
the direct sum of the $R_{V_i}$'s. So we have that
$$
\eqalign{
m_{\L\ba\del G}^{\tilde m_{\del \L},m_{\del G},
\eta_{\L\ba\del G},
\s_{\L\ba\del G}}\bigl |_{V_i}
&=R_{V_i}\left(
c m^*\s_{V_i}
+\frac{\eta_{V_i}}{q}
+\del_{V_i,\del V_i}m_{\del V_i}
+\del_{V_i,\del \L}\tilde m_{\del \L}
\right)\cr
&=: m_{V_i}^{\tilde m_{\del \L},m_{\del V_i},\eta_{V_i},
\s_{V_i}}
}
\tag{2.12}
$$
is a function depending only on what is appearing
as superscripts, namely random fields and Ising-spins {\it inside}
$V_{i}$ and continuous-spin boundary condition on $\del V_{i}$.
(The dependence on the global boundary condition 
$\tilde m_{\del \L}$
is of course only thru $\tilde m_{x}$ for $d(x,G_i)=1$.
We don't make this explicit in the notation.)

Also, the conditional fluctuation-Hamiltonian 
on $\L\ba\del G$ decomposes 
into a sum over connected components of its
support $\L\ba\del G$: 
$$
\eqalign{
&\D H^{\tilde m_{\del \L},m_{\del G},\eta_{\L\ba\del G},\s_{\L\ba\del G}}
_{\L\ba\del{G}}\left(m_{\L\ba\del{G}} \right)
=\sum_{i}\D H^{\tilde m_{\del \L},m_{\del V_i},
\eta_{V_i},\s_{V_i}}_{V_i}\left(m_{V_i} \right)\text{where}\cr
&\D H^{\tilde m_{\del \L},m_{\del V_i},
\eta_{V_i},\s_{V_i}}_{V_i}\left(m_{V_i} \right)\cr
&=\frac{1}{2}<\left(m_{V_i}
-m_{V_i}^{\tilde m_{\del \L},m_{\del V_i},\eta_{V_i},\s_{V_i}}
\right),
\left(a-q\D_{V_i}\right)\left(m_{V_i}-m_{V_i}^{\tilde m_{\del \L},m_{\del V_i},\eta_{V_i},\s_{V_i}}\right)>_{V_i}
}
\tag{2.13}
$$
Putting together the connected components of 
$\L\ba \ov{G}$ we can thus write
$$
\eqalign{
&\D H^{\tilde m_{\del \L},m_{\del G},
\eta_{\L\ba\del{G}},\s_{\L\ba\del{G}}}
_{\L\ba\del{G}}\left(m_{\L\ba\del{G}} \right)\cr
&=\D H^{\tilde m_{\del \L},m_{\del G},
\eta_{\L\ba\ov{G}},\s_{\L\ba\ov{G}}}
_{\L\ba\del{G}}\left(m_{\L\ba\ov{G}} \right)
+\sum_{{G_i}\atop{\hbox{ conn.cp of }G}}
\D H^{\tilde m_{\del \L},m_{\del G_i},
\eta_{G_i},\s_{G_i}}_{G_i}\left(m_{G_i} \right)\cr
}
\tag{2.14}
$$
So, the sum over $G$'s in (2.6) can be written as

$$
\eqalign{
&\sum_{G:\em \neq G \subset \L } 
\int_{\R^{\L}}dm_{\L}e^{
-H^{\tilde m_{\del \L},\eta_{\L}, \s_{\L}}_{\L}
\left(m_{\L}\right)}\cr
&\quad\times \prod_{{G_i}\atop{\hbox{ conn.cp of }G}}
\prod_{x\in \del G_i}1_{m_x\in U}\left[
\prod_{x\in G_i}\left(1_{m_x\not\in U}+w(m_x) \right)
-\prod_{x\in G_i}1_{m_x\not\in U}
\right]\cr
&=\sum_{G:\em \neq G \subset \L } 
e^{-\inf_{m'_{\L}} H^{\tilde m_{\del \L},\eta_{\L}, 
\s_\L}_{\L}\left(m'_{\L}\right)}\cr
&\int dm_{\del G}
e^{-\D H^{\tilde m_{\del \L},
\eta_{\L},\s_{\L}}_{\del G,\L}\left(m_{\del G} \right)}
\prod_{x\in \del G}1_{m_x\in U}
\int dm_{\L\ba\ov G}e^{
-\D H^{\tilde m_{\del \L},m_{\del G},
\eta_{\L\ba\ov{G}},\s_{\L\ba\ov{G}}}
_{\L\ba\ov{G}}\left(m_{\L\ba\ov{G}} \right)}\cr
&\quad\times \prod_{{G_i}\atop{\hbox{ conn.cp of }G}}
\int dm_{G_i}e^{-\D H^{\tilde m_{\del \L},m_{\del G_i},
\eta_{G_i},\s_{G_i}}_{G_i}\left(m_{G_i} \right)}
\left[
\prod_{x\in G_i}\left(1_{m_x\not\in U}+w(m_x) \right)
-\prod_{x\in G_i}1_{m_x\not\in U}
\right]\cr
}
\tag{2.15}
$$
Now we note the pleasant fact
that the Gaussian integral over $\L\ba\ov{G}$
is independent of all of the superindexed quantities
(since they appear only in the shift of the quadratic form),
so that it can be pulled out of the $m_{\del G}$-integral. 
It gives 
$$
\eqalign{
&\int dm_{\L\ba\ov G}e^{
-\D H^{\tilde m_{\del \L},m_{\del G},
\eta_{\L\ba\ov{G}},\s_{\L\ba\ov{G}}}
_{\L\ba\ov{G}}\left(m_{\L\ba\ov{G}} \right)}
=\left(2\pi \right)^{\frac{|\L-\ov{G}|}{2}}
\left(\det \left(a-q \D_{\L\ba\ov G} \right)\right)^{-\frac{1}{2}}
\cr
}
\tag{2.16}
$$
Let us look at the last line now. 
Conditional on $m_{\del G}$
we define anharmonic activities  by the formula
$$
\eqalign{
&I^{\tilde m_{\del \L},m_{\del G_i},
\eta_{G_i},\s_{G_i}}_{G_i}
:=\int dm_{G_i}e^{-\D H^{\tilde m_{\del \L},m_{\del G_i},
\eta_{G_i},\s_{G_i}}_{G_i}\left(m_{G_i} \right)}
\left[
\prod_{x\in G_i}\left(1_{m_x\not\in U}+w(m_x) \right)
-\prod_{x\in G_i}1_{m_x\not\in U}
\right]\cr
}
\tag{2.17}
$$
We write  $I^{\tilde m_{\del \L},m_{\del G},
\eta_{G},\s_{G}}_{G}=1$ for $G=\em$.
So we have obtained the following
representation for the non-normalized Ising-weights

\lemma{2.3}{\it With the above notations we have
$$
\eqalign{
&e^{b|\L|}Z_{\L}^{\tilde m_{\del \L}}(\s_{\L})
=
e^{-\inf_{m'_{\L}} H^{\tilde m_{\del \L},\eta_{\L}, 
\s_\L}_{\L}\left(m'_{\L}\right)}\cr
&\times\sum_{G:\em\sb G \subset \L } 
\left(2\pi \right)^{\frac{|\L\ba\ov{G}|}{2}}
\left(\det \left(a-q \D_{\L\ba\ov G} \right)\right)^{-\frac{1}{2}}\cr
&\int dm_{\del G}
e^{-\D H^{\tilde m_{\del \L},
\eta_{\L},\s_{\L}}_{\del G,\L}\left(m_{\del G} \right)}
\prod_{x\in \del G}1_{m_x\in U}
\prod_{{G_i}\atop{\hbox{ conn.cp of }G}}
I^{\tilde m_{\del \L},m_{\del G_i},
\eta_{G_i},\s_{G_i}}_{G_i}\cr
}
\tag{2.18}
$$
}

Let us pause for a minute and comment
on what we have obtained. 
For the purely Gaussian 
model (i.e. the $w$-terms are identically zero)
the contributions for $G\neq\em$ vanish.
So the above formula is a good starting
point for the derivation of the signed-contour
representation whose main contributions
are provided by the minimum of the 
Gaussian Hamiltonians in the first line.  
The main other non-trivial ingredient 
are the preliminary anharmonic activities 
$I^{\tilde m_{\del \L},m_{\del G},
\eta_{G},\s_{G}}_{G}$.
First of all, the whole construction 
makes only sense, if we are able to 
prove a suitable Peierls estimate for them,
to be discussed soon. They factorize 
over connected components $G_i$ of the
set $G$. The conditioning on $m_{\del G}$
has allowed us to have them {\it local} 
in the sense that they depend only on random fields 
and Ising-spins  {\it inside} $G_i$. 
Note that such a factorization does of course 
not hold for the remaining integral over $\del G$ 
(that would mean: over connected components of $\del G$),
as it is clear from (2.8).
Indeed, the fields $m_{\del G}$ fluctuate
according to the covariance matrix in the total volume $\L$.
So to speak, their (stochastic) dependence is mediated by 
the Gaussian local specification defined with (2.8).
Furthermore, the dependence of their mean-value 
in this local specification is 
(weakly) on {\it all} Ising-spins and random fields
in $\L$. Both kinds of dependence will have to be expanded later
in Chapter III when the integral
over $\del G$ is carried out.  
This will be done by enlarging the 
`polymers' $G$ and performing a 
high-temperature expansion.
Finally, the determinants provide only trivial 
modifications of the weights that we will obtain; 
they can easily be handled by a random walk expansion. 

Let us stress the following
nice feature of the above representation:
`Low-temperature contours' (see Chapter IV)
will be created only by the global energy-minimum
in the first line. Consequently 
there will be no complicated boundary terms 
for these `low-temperature' terms 
(that could be easily produced by a careless expansion).

Our further treatment of the expansion 
will be done under the assumption
of the following two properties:

\noindent{\thbf Positivity of anharmonic weights: }{\it
$$
\eqalign{
&I^{\tilde m_{\del \L},m_{\del G},
\eta_{G},\s_{G}}_{G}\geq 0
}
\tag{2.19}
$$
for all connected $G$, and $\tilde m_{\del \L}\in U^{\del \L}$,
$m_{\del G}\in U^{\del G}$, 
$\eta_{G}\in [-\d,\d]^G$,  $\s_{G}\in \{-1,1\}^G$.}

\noindent{\thbf Uniform Peierls Condition for anharmonic weights: }{\it
$$
\eqalign{
&I^{\tilde m_{\del \L},m_{\del G},
\eta_{G},\s_{G}}_{G}\leq  \e^{|G|}
}
\tag{2.20}
$$
for all connected $G$, and $\tilde m_{\del \L}\in U^{\del \L}$,
$m_{\del G}\in U^{\del G}$, 
$\eta_{G}\in [-\d,\d]^G$,  $\s_{G}\in \{-1,1\}^G$ with $\e>0$.
}

\bigskip

Rather than trying to be exhaustive in the description
of potentials that satisfy these conditions we will 
use the rest of this Chapter to fix
some properties that
imply them and discuss in detail the explicit
example of the $\phi^4$-theory in Lemma 2.6.
This should however indicate how the above
two conditions can be achieved in concrete cases by suitable
choices of the neighborhood $U$ and the constants
$a$ and $b$ occuring in the quadratic potential. 
The expansion will be continued in Chapter III.

Let us start by fixing the following almost 
trivial one-site
criterion. It makes sense if we 
are assuming the nearest neighbor coupling $q$ 
to be small.

\lemma{2.4}{\it
Suppose that $w(m_x)\geq 0$
for $m_x\in U$.

\item{(i)} Assume that we have uniformly 
for all choices of superindices
$$
\eqalign{
&\int dm_{x}e^{-\frac{a+4dq}{2}
\left( 
m_x- m^{\tilde m_{\del \L},m_{\del G},\eta_{G},\s_{G}}_x
\right)^2 }w(m_x)1_{m_x\in U}
\cr
&\geq \int dm_{x}e^{-\frac{a}{2}
\left( 
m_x- m^{\tilde m_{\del \L},m_{\del G},\eta_{G},\s_{G}}_x
\right)^2 }1_{m_x\not\in U}
}
\tag{2.21}
$$
Then we have the positivity (2.19).

\item{(ii)} Assume that 
$$
\eqalign{
&
\int dm_{x}e^{-\frac{a}{2}
\left( 
m_x- m^{\tilde m_{\del \L},m_{\del G},\eta_{G},\s_{G}}_x
\right)^2 }\left(w(m_x)1_{m_x\in U}+(1+w(m_x))1_{m_x\not\in U}
\right)\leq \e\cr
}
\tag{2.22}
$$
Then we have the uniform Peierls estimate (2.20)
with the same $\e$.

}

\proof Since we always have  $-1\leq w(m_x)<\infty$
the assumption $1_{m_x\in U}w(m_x)\geq 0$ implies that  
$$
\eqalign{
&\int dm_{G}e^{-\D H^{\tilde m_{\del \L},m_{\del G},
\eta_{G},\s_{G}}_{G}\left(m_{G} \right)}
\prod_{x\in G}\left(1_{m_x\not\in U}+w(m_x) \right)\cr
&\geq 
\int dm_{G}e^{-\D H^{\tilde m_{\del \L},m_{\del G},
\eta_{G},\s_{G}}_{G}\left(m_{G} \right)}
\prod_{x\in G}w(m_x)1_{m_x\in U}
\geq 0
\cr
}
\tag{2.23}
$$
We reduce the estimation of the integrals
to product integration by 
the pointwise estimate on the quadratic form
$$
\eqalign{
& a \Vert v_G \Vert_2^2\leq 
<v_{G},\left(a-q\D^{D}_{G}\right)v_G>_{G}
\leq (a+4d q) \Vert v_G \Vert_2^2
\cr
}
\tag{2.24}
$$ 
This gives 
$$
\eqalign{
&\int dm_{G}e^{-\D H^{\tilde m_{\del \L},m_{\del G},
\eta_{G},\s_{G}}_{G}\left(m_{G} \right)}
\prod_{x\in G}w(m_x)1_{m_x\in U}\cr
&\geq 
\prod_{x\in G}
\int dm_{x}e^{-\frac{a+4dq}{2}
\left( 
m_x- m^{\tilde m_{\del \L},m_{\del G},\eta_{G},\s_{G}}_x
\right)^2 }w(m_x)1_{m_x\in U}
\cr
}
\tag{2.25}
$$
and, on the other hand,  
$$
\eqalign{
&
\int dm_{G}e^{-\D H^{\tilde m_{\del \L},m_{\del G},
\eta_{G},\s_{G}}_{G}\left(m_{G} \right)}
\prod_{x\in G}1_{m_x\not\in U}\cr
&\leq \prod_{x\in G}
\int dm_{x}e^{-\frac{a}{2}
\left( 
m_x- m^{\tilde m_{\del \L},m_{\del G},\eta_{G},\s_{G}}_x
\right)^2 }1_{m_x\not\in U}\cr
}
\tag{2.26}
$$
This proves (i).

The Peierls estimate (ii) follows 
from dropping the second product in the 
definition of $I$ and using (2.24) to write  
$$
\eqalign{
&I^{\tilde m_{\del \L},m_{\del G},
\eta_{G},\s_{G}}_{G}\cr
&\leq
\int dm_{G}e^{-\D H^{\tilde m_{\del \L},m_{\del G},
\eta_{G},\s_{G}}_{G}\left(m_{G} \right)}
\prod_{x\in G}\left((1+w(m_x))1_{m_x\not\in U}+w(m_x)1_{m_x\in U}
\right)\cr
&\leq \prod_{x\in G}
\int dm_{x}e^{-\frac{a}{2}
\left( 
m_x- m^{\tilde m_{\del \L},m_{\del G},\eta_{G},\s_{G}}_x
\right)^2 }\left((1+w(m_x))1_{m_x\not\in U}+w(m_x)1_{m_x\in U}
\right)\cr
}
\tag{2.27}
$$
\endproof

\bigskip\bigskip

Next we compute how big the nearest neighbor coupling
$q$ and size of the random fields $\d$
can be in order that any boundary condition
in $U$ yields a minimizer of the Gaussian Hamiltonian
on $G$ that is `well inside' $U$.
We have

\lemma{2.5}{\it Let $0<A_1\leq A_2$ and
$U^+=[m^*-A_2, m^*+A_2]$, $U=U^+\cup \left(-U^+\right)$.
Assume that 
$q\leq \frac{a}{2d}\left(  
\frac{2 m^*+ A_2}{A_1}-1
\right)^{-1}
$
and $\d\leq \frac{a A_1}{2}$.
Then we have that 
$$
\eqalign{
\left|m_{x}^{\tilde m_{\del \L},m_{\del G},\eta_{G},\s_{G}}
-m^* \s_x
\right|\leq A_1
}
\tag{2.28}
$$
for all $G$, $\tilde m_{\del \L}\in U^{\del \L}$,
$m_{\del G}\in U^{\del G}$, 
$\eta_{G}\in [-\d,\d]^G$,  $\s_{G}\in \{-1,1\}^G$.}

\proof 
Note the linear dependence 
$m_{x}^{\tilde m_{\del \L},m_{\del G},\eta_{G},\s_{G}}
=m_{x}^{\tilde m_{\del \L},m_{\del G},\eta_{G}=0,\s_{G}}
+\left(R_{G}\frac{\eta_{G}}{q}\right)_x
$.
Let us thus choose the condition for $q$ 
s.t. 
$$
\eqalign{
\left|m_{x}^{\tilde m_{\del \L},m_{\del G},\eta_{G}=0,\s_{G}}
-m^* \s_x
\right|\leq \frac{A_1}{2}  
}
\tag{2.29}
$$
This condition is in fact achieved for a one-point $G=\{x\}$
and the boundary conditions having the `wrong sign'
with modulus $m^*+A_2$ as we will formally see as follows. 
Let us assume that $\s_x=-1$ and write this time
for simplicity $\del G$ for the boundary 
in $\Z^d$ (including possible 
sites in the outer boundary of $\L$ in $\Z^d$).
Then we have, due to the positivity of the matrix elements
of $R_{G}$ that 
$$
\eqalign{
&m_{x}^{\tilde m_{\del \L},m_{\del G},\eta_{G}=0,\s_{G}}\cr
&\leq -R_{G;x,x} c m^*
+\sum_{y\in G\ba\{x\}}R_{G;x,y} c m^*
+\left(R_{G}\del_{G,\del G}1_{\del G}(m^* +A_2)\right)_{x}\cr
}
\tag{2.30}
$$
We employ the equation
$R_{G}\left( c 1_{G}
+\del_{G,\del G} 1_{\del G}\right)= 1_{G}$
to  write the last line of (2.30) as
$$
\eqalign{
&m^* -2 R_{G;x,x} c m^*
+ A_2 - A_2\left(R_{G}c 1_{G}\right)_{x}
}
\tag{2.31}
$$
We note that $R_{G;x,x}$ is an increasing function 
in the sets $G\ni x$ (which can be seen by the random walk
representation, see Appendix (A.8)).
Further $\left(R_{G}1_{G}\right)_{x}$ is an increasing function in $G$.
So the maximum over $G$ of (2.31) is achieved 
for $G=\{x\}$. With $R_{\{x\};x,x}=\frac{1}{c+2d}$
the value of (2.31) becomes
$-m^* +\left(2 m^*
+ A_2\right)\frac{2d}{c+2d} 
$
which gives the upper bound
$$
\eqalign{
m_{x}^{\tilde m_{\del \L},m_{\del G},\eta_{G}=0,\s_{G\ba\{x\}},\s_x=-1}
+m^* \leq \left(2 m^*
+ A_2\right)\frac{2d}{c+2d} 
}
\tag{2.32}
$$
In the same way we obtain
$$
\eqalign{
m_{x}^{\tilde m_{\del \L},m_{\del G},\eta_{G}=0,\s_{G\ba\{x\}},\s_x=-1}
+m^* \geq -A_2\frac{2d}{c+2d} 
}
\tag{2.33}
$$
Equating of the r.h.s. with $A_1/2$ gives the r.h.s. of 
the condition on $q$ stated in the hypothesis.

For the estimate of the random field term  
note that $0\leq R_{G;x,y}\leq R_{Z^d;x,y}$
and $\sum_{y\in \Z^d}R_{Z^d;x,y}=\frac{1}{c}$
which give us 
$$
\eqalign{
\left|\sum_{y\in G}R_{G;x,y}\frac{\eta_{y}}{q}\right|
\leq \frac{\d}{q} \sum_{y\in \Z^d}
R_{Z^d;x,y}=\frac{\d}{a}\leq \frac{A_1}{2}
}
\tag{2.34}
$$
\endproof

At this stage the treatment 
has to be made specific to the 
concrete potential and we specialize to our
example, the $\phi^4$-theory with potentials given by (1.2).
The following Lemma summarizes how we can produce
positivity and 
an arbritrarily small anharmonic Peierls constant.
More specific information can be found in the proof.

\lemma{2.6}{\it 
For fixed $\e_0>0$ we put
$$
\eqalign{
&U^+=[m^*-\left(\e_0 {m^*}\right)^\frac{1}{3},m^*+
\left(\e_0 {m^*}\right)^\frac{1}{3}]
}
\tag{2.35}
$$
Then we have 

\item{(i)} For any value of $\e_0,m^*,q,\d$
there exists a choice of parameters $a$ and $b$
such that the anharmonic weights obey the positivity (2.19).

\noindent Furthermore there exist strictly positive 
constants $a(m^*,\e_0)$, $b(m^*,\e_0)$, $q_0(m^*,\e_0)$, and 
$\d_0(m^*,\e_0)$ such that the following is true. 

\item{(ii)} For all $q\leq q_0(m^*,\e_0)$ and  $\d\leq \d_0(m^*,\e_0)$
we have the  Peierls estimate 
(2.20) with a constant $\e(\e_0,m^*)$ that is
independent of $q,\d$.  

\item{(iii)} If $\e_0$ is small enough
this constant obeys the estimate  
$\e(\e_0,m^*)\leq \frac{\e_0}{10}$ 
whenever $m^*\geq m^*_0(\e_0)$ is large enough.

\noindent The above constants can be chosen like 
$$
\eqalign{
&a(m^*,\e_0)=\frac{(2+{\e_0}^{\frac{1}{3}}{m^*}^{-\frac{2}{3}})^2}{4}
\sim 1 \cr
&q(m^*,\e_0)=
\frac{a(m^*,\e_0)}{2d}\left(20 {\e_0}^{-\frac{1}{3}}
{m^*}^{\frac{2}{3}}  +9\right)^{-1}, \quad
\d_0(m^*,\e_0)= \frac{a(m^*,\e_0)
\left(\e_0 {m^*}\right)^\frac{1}{3}}{20}
}
\tag{2.36}
$$
and $b(m^*,\e_0)\sim e^{-\const {m^*}^{\frac{2}{3}}}$  
with $m^{*}\uparrow \infty$.
}

\proof We will take time to 
motivate our choices of the parameters
that are made to ensure the validity of the
assumptions of Lemma 2.4.
Let us write the neighborhood $U^+$ in the form
$U^+=[(1-\e_1)m^*,(1+\e_1)m^*]$
and show why the choice of $\e_1$ given in 
(2.35) comes up. 
The zeroth requirement 
on $a$ and $b$ we have to meet is $w(m_x)1_{m_x\in U}\geq 0$.
So, let us choose the Gaussian curvature $a>1$
to be the smallest number s.t. we have, for all $m_x\in U^+$, that
the Gaussian centered around $m^*$ is dominated by the true
potential i.e. 
$$
\eqalign{
&e^{-\frac{a\left(m_x-m^*\right)^2}{2}}\leq e^{-V(m_x)}
}
\tag{2.37}
$$
with equality for $m_x=(1+\e_1)m^*$.
This amounts to
$a=\frac{(2+\e_1)^2}{4}$,
as in (2.36).  Then we have on $U^+$ 
for the Gaussian centered around $-m^*$
$$
\eqalign{
&e^{-\frac{a\left(m_x+m^*\right)^2}{2}}
\leq 
e^{-
\frac{(2+\e_1)^2 (2-\e_1)^2+1-(1+\e_1)^2}{8}{m^*}^2}e^{-V(m_x)}
}
\tag{2.38}
$$
which gives us the estimate
$$
\eqalign{
&1+w(m_x)
\geq e^{b}
\left[
1+e^{-
\frac{(2+\e_1)^2 (2-\e_1)^2+1-(1+\e_1)^2}{8}{m^*}^2}
\right]^{-1}
}
\tag{2.39}
$$
on $U^+$. Any choice of $e^b$ bigger than the denominator 
thus ensures $w(m_x)1_{m_x\in U}\geq 0$.

To have property (i) in Lemma 2.4. 
we have to choose $e^b$ even bigger.
Obviously it is implied by
$$
\eqalign{
&\inf_{m_x\in U^+}w(m_x)
\geq  \frac{\int dm_{x}e^{-\frac{a}{2}
\left( 
m_x- m^{\tilde m_{\del \L},m_{\del G},\eta_{G},\s_{G}}_x
\right)^2 }1_{m_x\not\in U}}
{\int dm_{x}e^{-\frac{a+4dq}{2}
\left( 
m_x- m^{\tilde m_{\del \L},m_{\del G},\eta_{G},\s_{G}}_x
\right)^2 }1_{m_x\in U}
}
}
\tag{2.40}
$$
But note that we always have 
$$
\eqalign{
&\left| 
\left|m^{\tilde m_{\del \L},m_{\del G},\eta_{G},\s_{G}}_x\right|
-m^*
\right|\leq \hat m^{\hbox{max}}\left(m^*,\d,q,a\right)
}
\tag{2.41}
$$
with a constant $\hat m^{\hbox{max}}\left(m^*,\d,q,a\right)$
that is finite for any fixed $m^*,\d,q,a$ and that
is estimated by Lemma 2.5. So the trivial choice 
$$
\eqalign{
&e^{b\left(m^*,\d,q,a\right)}
:=\left(
1+e^{-
\frac{(2+\e_1)^2 (2-\e_1)^2+1-(1+\e_1)^2}{8}{m^*}^2}
\right)\cr
&\times \left(
1+
\sup_{\hat m:|\hat m|\leq  \hat m^{\hbox{\srm max}}\left(m^*,\d,q,a\right)}
\frac{\int dm_{x}e^{-\frac{a}{2}
\left( 
m_x- \hat m \right)^2 }1_{m_x\not\in U}}
{\int dm_{x}e^{-\frac{a+4dq}{2}
\left( 
m_x- \hat m
\right)^2 }1_{m_x\in U}
}\right)
}
\tag{2.42}
$$
gives some finite number and
ensures the positivity of the anharmonic activities. 
This proves (i).

Let us now turn to quantitative estimates
on the Peierls constant.
To start with, the above definition of $b$ is of course only
useful if $b$ will be small. 
Now, the r.h.s. of (2.42) is small whenever 
the centering of the Gaussian integrals
is `safe' inside $U$ and the neighborhood 
$U$ is big enough to carry most of the Gaussian integral. 
We apply Lemma 2.5. with $A_2=\e_1 m^*$ and
$A_1=\frac{A_2}{10}$. The hypotheses of the Lemma 
then give us the conditions 
$q\leq q_0$ and $\d\leq \d_0$
with 
$$
\eqalign{
&q_0= \frac{a}{2d}\left(\frac{20}{\e_1}+9\right)^{-1},
\qquad\d_0= \frac{a\e_1 m^*}{20}
}
\tag{2.43}
$$
Then we have
$$
\eqalign{
&\frac{\int dm_{x}e^{-\frac{a}{2}
\left( 
m_x- m^{\tilde m_{\del \L},m_{\del G},\eta_{G},\s_{G}}_x
\right)^2 }1_{m_x\not\in U}}
{\int dm_{x}e^{-\frac{a+4dq}{2}
\left( 
m_x- m^{\tilde m_{\del \L},m_{\del G},\eta_{G},\s_{G}}_x
\right)^2 }1_{m_x\in U}
}
\leq \sqrt\frac{a+4dq}{a}
\frac{
\P\left[
|G|\geq \sqrt{a}\frac{9\e_1 {m^*}}{10}\right]
}
{1-\P\left[
|G|\geq \sqrt{a+ 2dq}\frac{9\e_1 {m^*}}{10}\right]
}
}
\tag{2.44}
$$
This shows that $b\sim e^{-\const\,\cdot\, (\e_1 m^* )^2}$ 
tends to zero rapidly
if $\e_1 m^*$ is getting large.

Let us now see what Peierls constant
we get according to Lemma 2.4 (ii).
This will explain why the neighborhood $U^+$
should in fact be of the form (2.35).

Our choice of $U$ and $a$ yields
that we have, for all $m_x\in U^+$, that
$$
\eqalign{
&e^{-V(m_x)+\frac{a\left(m_x-m^*\right)^2}{2}}
\leq e^{\e_1\left(m_x-m^*\right)^2}
}
\tag{2.45}
$$
This gives
$1+w(m_x)\leq e^{b+\e_1\left(m_x-m^*\right)^2}$.
From this we have
$$
\eqalign{
&\int dm_{x}e^{-\frac{a}{2}
\left( 
m_x- m^{\tilde m_{\del \L},m_{\del G},\eta_{G},\s_{G}}_x
\right)^2 }\left(1+w(m_x)\right)1_{m_x\in U^{\s_x}}\cr
&\leq  e^{b} 
\int dm_{x}e^{-\frac{a}{2}
\left( 
m_x- m^{\tilde m_{\del \L},m_{\del G},\eta_{G},\s_{G}}_x
\right)^2 +\e_1\left(m_x-m^*\right)^2}\cr
&= e^{b}\sqrt{\frac{2\pi}{a-2\e_1}}e^{\frac{a\e_1}{a-2\e_1}\left( 
m^{\tilde m_{\del \L},m_{\del G},\eta_{G},\s_{G}}_x-m^*
\right)^2}
\leq  
e^{b}\sqrt{\frac{2\pi}{a-2\e_1}}e^{\frac{a\e_1^3 {m^*}^2}{100(a-2\e_1)}
}
\cr
}
\tag{2.46}
$$
and hence 
$$
\eqalign{
&\int dm_{x}e^{-\frac{a}{2}
\left( 
m_x- m^{\tilde m_{\del \L},m_{\del G},\eta_{G},\s_{G}}_x
\right)^2 }\left(1+w(m_x)\right)1_{m_x\in U}\cr
&\leq 2 e^{b}
\Bigl[
\sqrt{\frac{2\pi}{a-2\e_1}}e^{\frac{a\e_1^3 {m^*}^2}{100(a-2\e_1)}}
-\sqrt{\frac{2\pi}{a}}
\P\left[
|G|\leq \sqrt{a}\frac{9 \e_1 m^*}{10}
\right]
\Bigr]
}
\tag{2.47}
$$
Indeed, the l.h.s. is $\OO(\e_1^3 {m^*}^2)+\OO(\e_1)$
and thus imposes the condition 
that $\e_1^3 {m^*}^2$ be small! 
This estimate can essentially
not be improved upon. It determines 
the dependence  of the Peierls constant $\e$ on $\e_1$
and $m^*$.

Finally, the integrals over $U^c$ are much smaller: 
Indeed, for the bounded part of $U^c$ we estimate
$$
\eqalign{
&
\int_0^{(1-\e_1)m^*} dm_{x}e^{-\frac{a}{2}
\left( 
m_x- m^{\tilde m_{\del \L},m_{\del G},\eta_{G},\s_{G}}_x
\right)^2 }(1+w(m_x))\cr
&
\leq \int_0^{(1-\e_1)m^*} dm_{x}e^{-\frac{a}{2}
\left( 
m_x- m^{\tilde m_{\del \L},m_{\del G},\eta_{G},\s_{G}}_x
\right)^2 }e^{-V(m_x)+\frac{a\left(m_x-m^*\right)^2}{2}}
\cr
&
= e^{-\frac{a}{2}
\left( 
m^*- m^{\tilde m_{\del \L},m_{\del G},\eta_{G},\s_{G}}_x
\right)^2}  
\int_0^{(1-\e_1)m^*} dm_{x}e^{-a 
\left( 
m^*- m^{\tilde m_{\del \L},m_{\del G},\eta_{G},\s_{G}}_x
\right)\left(m_x-m^*\right)}e^{-V(m_x)}
\cr
}
\tag{2.48}
$$
We have for the last integral
$$
\eqalign{
&\int_0^{(1-\e_1)m^*} dm_{x}e^{-a 
\left( 
m^*- m^{\tilde m_{\del \L},m_{\del G},\eta_{G},\s_{G}}_x
\right)\left(m_x-m^*\right)}e^{-V(m_x)}
\cr
&\leq \int_0^{(1-\e_1)m^*} dm_{x}e^{\frac{a \e_1 m^*}{10} 
\left(m_x-m^*\right)}e^{-V(m_x)}\cr
&\leq \int_0^{(1-\e_1)m^*} dm_{x}e^{\frac{a \e_1 m^*}{10} 
\left(m_x-m^*\right)}e^{- \frac{\left(m_x-m^*\right)^2}{8}}\cr
}
\tag{2.49}
$$
The maximizer of the last exponent is $m_x=m^*+ \frac{ 2 a \e_1 m^{*}}{10}$
which is outside the range of integration (due to our choice
of the $10$ before (2.43))
Estimating for simplicity 
the integral by the value of the integrand
at $(1-\e_1)m^*$ just gives 
$$
\eqalign{
&
\int_0^{(1-\e_1)m^*} dm_{x}e^{-\frac{a}{2}
\left( 
m_x- m^{\tilde m_{\del \L},m_{\del G},\eta_{G},\s_{G}}_x
\right)^2 }(1+w(m_x))
\leq m^*  e^{-\left(
\frac{1}{8}-\frac{a}{10}
\right)
\left(\e_1 m^* \right)^2
}
\cr
}
\tag{2.50}
$$
For the unbounded part of 
of $U^c$ where
$m\geq m^*(1+\e_1)$ we have with our choice 
of $a$ that $1+w(m_x)\leq 1$.
This gives us 
$$
\eqalign{
&
\int_{(1+\e_1)m^*}^{\infty} dm_{x}e^{-\frac{a}{2}
\left( 
m_x- m^{\tilde m_{\del \L},m_{\del G},\eta_{G},\s_{G}}_x
\right)^2 }(1+w(m_x))\cr
&\leq \sqrt{\frac{2\pi}{a}}\P\left[
G\geq \sqrt{a}\frac{9 \e_1 m^*}{10}
\right]\leq e^{-\const \left(\e_1 m^*\right)^2}\cr
}
\tag{2.51}
$$
Collecting the terms gives our final estimate 
on the Peierls constant
$$
\eqalign{
&\e\leq 2 e^{b}
\Bigl[
\sqrt{\frac{2\pi}{a-2\e_1}}e^{\frac{a\e_1^3 {m^*}^2}{100(a-2\e_1)}}
-\sqrt{\frac{2\pi}{a}}\cr
&+ m^*  e^{-\left(
\frac{1}{8}-\frac{a}{10}
\right)
\left(\e_1 m^* \right)^2}
+3\sqrt{\frac{2\pi}{a}}\P\left[
G\geq \sqrt{a}\frac{9 \e_1 m^*}{10}
\right]
\Bigr]\cr
}
\tag{2.52}
$$
From here the lemma follows.
\endproof



\bigskip\bigskip

We are still due the 

\proofof{Lemma 2.1}
We expand 
$\prod_{x\in \L}\left(1 +w_x\right)
=1+\sum_{\L_0:\em\neq \L_0\sb \L}\prod_{x\in \L_0}w_x$.
Let $A(\L_0)\sb (\L\ba\UU)\ba \L_0$ denote the maximal
set amongst the sets $A\sb (\L\ba\UU)\ba \L_0$ that 
are connected to $\L_0$.
(We say that 
a set $A$ is connected to a set $\L_0$
iff, for each point $u$ in $A$,
there exists a nearest neighbor
path inside $A\cup \L_0$ that joins $u$ and 
some point in $\L_0$.) Equivalently, this $A(\L_0)$ is the 
unique set $A\sb \L\ba \L_0$ s.t. 
$x\not\in \UU$ for all $x\in A$ and 
$x\in \UU$ for all $x\in \del (\L_0\cup A)$.

We collect terms according to the sets $G=\L_0\cup A(\L_0)$.
Denoting by $G_i$
the connected components of $G$ and by $L_i=\L_0\cap G_i$ 
we have then
$$
\eqalign{
\prod_{x\in \L}\left(1+w_x\right)
&=1+\sum_{\L_0:\em\neq \L_0\sb \L}
\prod_{x\in A(\L_0)}1_{x\not\in \UU} 
\prod_{x\in \del (\L_0\cup A(\L_0))}1_{x\in \UU}
\prod_{x\in \L_0}w_x\cr
&=1+\sum_{G:\em \neq G \subset \L } 
\prod_{{G_i}\atop{\hbox{ conn.cp of }G}}
\sum_{L_i:\em\neq L_i\subset G_i} \prod_{x\in G_i\ba L_i}
1_{x\not\in \UU}
\prod_{x\in \del G_i}1_{x\in \UU}
\prod_{x\in L_i}
w_x
}
\tag{2.53}
$$
Adding and subtracting the term for $L_i=\em$ we have
$$
\eqalign{
&\sum_{L_i:\em\neq L_i\subset G_i} \prod_{x\in G_i\ba L_i}
1_{x\not\in \UU}\prod_{x\in L_i}w_x=
\prod_{x\in G_i}\left(1_{x\not\in \UU}+w_x \right)
-\prod_{x\in G_i}1_{x\not\in \UU}
}
\tag{2.54}
$$
which proves the lemma.\endproof


\overfullrule=0pt
\bigskip\bigskip
\chap{III. Control of Anharmonicity}

We start from the representation 
of Lemma 2.3 for 
the non-normalized Ising weights.
We assume positivity and Peierls condition
for the anharmonic ($I$-) weights 
as discussed in  Chapter II and verified
for the $\phi^4$-potential. 
Carrying out 
the last remaining continuous spin-integral
we express the last line in (2.18) 
in terms of activities that are positive,
obey a Peierls estimate and depend
in a {\it local way} on the Ising-spin 
configuration $\s_{\L}$
and the realization of the random fields $\eta_{\L}$.
We stress that all estimates that follow will be uniform
in the Ising-spin configuration and 
the configuration of the random field. 

The result of this is

\proposition{3.1}{\it 
Assume that the anharmonic $I$-weights (2.17)
satisfy the Positivity (2.19) 
and the uniform Peierls Condition
(2.20) with a constant $\e$. Suppose that $\e$
is sufficiently small, $q$ is sufficiently small, 
$a$ is of the order one,
$q(m^*)^2$ sufficiently large. 
Suppose that $\d\leq \Const m^*$ 
and $|U|\leq \Const m^*$
with constants of the order unity.   

Then, for any continuous-spin
boundary condition $\tilde m_{\del \L}\in U^{\del \L}$
and any realization of the random fields 
$\eta_{\L}\in [-\d,\d]^{\L}$,
the non-normalized Ising weights (2.1)
have the representation
$$
\eqalign{
&Z_{\L}^{\tilde m_{\del \L},\eta_{\L}}(\s_{\L})
=e^{-b|\L|}\left(2\pi \right)^{\frac{|\L|}{2}}
\left(\det \left(a-q \D_{\L} \right)\right)^{-\frac{1}{2}}
e^{-\inf_{m'_{\L}} H^{\tilde m_{\del \L},\eta_{\L}, 
\s_\L}_{\L}\left(m'_{\L}\right)}\cr
&\times\sum_{G:\em\sb G \subset \L } 
\bar\r^{\tilde m_{\del_{\del \L} G}}\left(G; \s_{G}
,\eta_{G} \right)\cr
}
\tag{3.1}
$$
where the activity $\bar\r$ appearing under the $G$-sum
is non-negative
and depends only on the indicated arguments. 
$\bar\r$ factorizes over the
connected components $G_i$ of its support $G$, i.e.
$$
\eqalign{
&\bar\r^{\tilde m_{\del_{\del \L} G}}
\left(G; \s_{G},\eta_{G} \right)
=\prod_{i}\bar\r^{\tilde m_{\del_{\del \L} G_i}}
\left(G_i; \s_{G_i},\eta_{G_i} \right)
\cr
}
\tag{3.2}
$$
and we have 
$\bar\r^{\tilde m_{\del_{\del \L} G}}
\left(G=\em; \s_{G},\eta_{G} \right)=1$.

\noindent $\bar\r$ has the `infinite volume 
symmetries'  of: 

\item{(a)} Invariance under joint flips of spins and random fields 
$\bar\r
\left(G; \s_{G},\eta_{G} \right)
=\bar\r\left(G; -\s_{G},-\eta_{G} \right)$
if $G$ does not touch the boundary (i.e. 
$\del_{\del \L} G=\em$) 

\item{(b)} Invariance under lattice shifts
$\bar\r
\left(G; \s_{G},\eta_{G} \right)
=\bar\r\left(G+t; \s_{G+t},\eta_{G+t} \right)$
if $G,G+t\sb \L$ don't touch the boundary

We have the uniform Peierls estimate
$$
\eqalign{
&\bar\r^{\tilde m_{\del_{\del \L} \bar G}}
\left(\bar G; \s_{\bar G},\eta_{\bar G} \right)
\leq  e^{-\a|\bar G|}
\cr
}
\tag{3.3}
$$
with $\a=\const \times \min\left\{
\log\frac{1}{q},\log\frac{1}{\e} \left(\frac{\log \frac{1}{q}}{\log m^*}
\right)^d\right\}$.
}

\noindent{\bf Remark 1:}
Note that the  first line of (3.1) gives 
the value  for vanishing anharmonicity (i.e. $w(m_x)\equiv 0$).

\noindent{\bf Remark 2:}
For any fixed Ising-spin $\s_{\L}$ and realization of random 
fields $\eta_{\L}$ 
the sum in the last line is the partition function of  
a non-translation invariant 
polymer model for polymers $G$.
Note that there is no suppression 
of the activities $\bar \r$ in the above bounds
in terms of the Ising-spins. 
From the point of view
of the polymers $G$ the Ising spins and random fields 
play the similar role of describing an 
`external disorder.'

\proofof{Proposition 3.1}
To yield this representation 
we must treat the last line of (2.18).
We can not carry out the $m_{\del G}$-integral
directly but need some further preparation
that allows us to treat the `long range' parts
of the exponent by a high-temperature expansion. 
Depending on the parameters of the model
(to be discussed below) we will then have to enlarge and 
glue together connected components of the 
support $G$ . 
For any set $G\sb \L$ we write
$$
\eqalign{
&G^r=\{x\in \L; d(x,G)\leq r\}
}
\tag{3.4}
$$
for the $r$-hull of $G$ in $\L$.
Then we have, under the assumptions 
on the parameters as in Proposition 3.1.

\lemma{3.2}{\it 
There is a choice of 
$r\sim \Const \frac{\log m^*}{\log\left(\frac{1}{q}  \right)}$
such that the following is true. 
For each fixed subset $G\sb \L$, continuous-spin
boundary condition $\tilde m_{\del \L}\in U^{\del \L}$, 
fixed  Ising-configuration $\s_{\L}\in \{-1,1\}^{\L}$ and 
random fields 
$\eta_{\L}\in [-\d,\d]^{\L}$
we can write
$$
\eqalign{
&\int dm_{\del G}
e^{-\D H^{\tilde m_{\del \L},
\eta_{\L},\s_{\L}}_{\del G,\L}\left(m_{\del G} \right)}\,\,
1_{m_{\del G}\in U^{\del G}}\,\,
I^{\tilde m_{\del \L},m_{\del G},
\eta_{G},\s_{G}}_{G}\cr
&=\left(2 \pi\right)^{\frac{|\del G|}{2}}
\sqrt{\det \left(\Pi_{\del G}\left(a-q\D_{G^r} \right)^{-1}\Pi_{\del G}\right)}
\sum_{{\tilde G:\tilde G\sb \L}\atop{G^r\sb \tilde G}}
\r^{\tilde m_{\del_{\del \L} \tilde G}}\left(G,\tilde G; \s_{\tilde G},\eta_{\tilde G} \right)\cr
}
\tag{3.5}
$$
where the activity appearing under the $\tilde G$-sum
depends only on the indicated arguments and 
obeys the uniform bounds
$$
\eqalign{
&0\leq \r^{\tilde m_{\del_{\del \L} \tilde G}}
\left(G,\tilde G; \s_{\tilde G},\eta_{\tilde G} \right)
\leq  e^{-\bar\a|\tilde G|}
\cr
}
\tag{3.6}
$$
with $\bar\a=\Const \times \min\left\{
\log\frac{1}{q},\log\frac{1}{\e} \left(\frac{\log \frac{1}{q}}{\log m^*}
\right)^d\right\}$.
It factorizes over the
connected components $\tilde G_i$ of the set $\tilde G$, i.e.
$$
\eqalign{
&\r^{\tilde m_{\del_{\del \L} \tilde G}}
\left(G,\tilde G; \s_{\tilde G},\eta_{\tilde G} \right)
=\prod_{i}\r^{\tilde m_{\del_{\del \L} \tilde G_i}}
\left(G\cap \tilde G_i,
\tilde G_i; \s_{\tilde G_i},\eta_{\tilde G_i} \right)
\cr
}
\tag{3.7}
$$
For $\tilde G$ not touching the boundary (i.e. 
$\del_{\del \L} \tilde G=\em$) $\r$
is invariant under joint flips of spins and random fields
and lattice shifts.
}

\remark Later it will be convenient to have the 
determinant appearing on the
r.h.s.; in fact it could also 
be absorbed in the activities under the $\tilde G$-sum.

\proofof{Lemma 3.2}
Let us recall definition (2.8) of the `fluctuation-
Hamiltonian'  
(involving the global minimizer (2.10)) which gives
the Hamiltonian of the projection onto $\del G$ of an 
Ising-spin and random-field dependent Gaussian field in $\L$.
Our first step is to decompose
this projection from $\L$ onto $\del G$
into a `low temperature-part' and a `high temperature-part'.
For fixed $G$ we 
will consider definition (2.8) where $\L$ will 
be replaced by $G^r$; for $r$ large enough 
the resulting term `low-temperature'- term
is close enough to the full 
expression, so that the rest can be treated 
by a high-temperature expansion. 

We write $\del_{B}:=\{x\in B; d(x,A)=1\}$
for the outer boundary in a set $B\sb \Z^d$. Recall that, with 
this notation $\del A=\del_{\L}A$, 
so that $\del_{\Z^d}(G^r)=
\del_{\del \L}(G^r)\cup \del (G^r)$.

Then the  precise form of the decomposition
we will use reads

\lemma{3.3}{\it With a suitable choice of 
$r\sim \Const \frac{\log m^*}{\log\left(\frac{1}{q}  \right)}$
we have
$$
\eqalign{
&
\D H^{\tilde m_{\del \L},
\eta_{\L},\s_{\L}}_{\del G,\L}\left(m_{\del G} \right)\cr
&=
\D H^{\left(\tilde m_{\del_{\del \L} G^r},
0_{\del_{\L} G^r}\right),
\eta_{G^r},\s_{G^r}}_{\del G,G^r}\left(m_{\del G} \right)
+\sum_{{C \sb \L}\atop
{C\cap\del G\neq\em;    C \cap(G^r)^c\neq \em}} 
\bar H^{\hbox{\srm HT}}_{\del G,G^r }(m_{\del G},
\s_{G^r},\eta_{G^r}
;C, \s_{C},\eta_{C})\cr
}
\tag{3.8}
$$
where the functions appearing under the $C$-sum
depend only on the indicated arguments and 
obey the uniform bound
$$
\eqalign{
&\left|\bar H^{\hbox{\srm HT}}_{\del G,G^r }(m_{\del G},
\s_{G^r},\eta_{G^r}
;C, \s_{C},\eta_{C})\right|
\leq 
e^{-\tilde \a|C|}
\cr
}
\tag{3.9}
$$
uniformly  in $m_{\del G}\in U^{\del G}$
and all other quantities 
for the $C$'s occuring in the sum in (3.8).
Here $\tilde \a =\const \log\frac{1}{q}$.
}

\remark Note that the first part (`low temperature-part')
decomposes of course 
over the connected components $(G^r)_i$ of $G^r$, i.e. 
$$
\eqalign{
&\D H^{\left(\tilde m_{\del_{\del \L} G^r},
0_{\del_{\L} G^r}\right),
\eta_{G^r},\s_{G^r}}_{\del G,G^r}\left(m_{\del G} \right)
=\sum_{i}
\D H^{\left(\tilde m_{\del_{\del \L} (G^r)_i},
0_{\del_{\L} (G^r)_i}\right),
\eta_{(G^r)_i},\s_{(G^r)_i}}_{\del G\cap (G^r)_i,(G^r)_i}
\left(m_{\del G\cap (G^r)_i} \right)\cr
}
\tag{3.10}
$$

\proofof{Lemma 3.3} The l.h.s. and the first term
on the r.h.s. of (3.8) differ in two places:
The matrix and the centerings.  
We expand both differences using 
the random walk representation.

The decomposition of the matrix 
into the matrix where $\L$ is replaced 
by $G^r$ and a remainder term can be written as
$$
\eqalign{
&\left(\Pi_{\del G}R_{\L}\Pi_{\del G}\right)^{-1}\cr
&=\left(\Pi_{\del G}R_{G^r}\Pi_{\del G}\right)^{-1}
-\sum_{{C \sb \L\ba\del G}\atop
{C\cap(G^r)^c\neq \em,{C\cap G^2}\neq \em }}
\del_{\del{G},\L\ba\del{G}}\RR\left(\cdot\,\rightarrow\cdot\,; C\right)
\del_{\L\ba\del{G},\del{G}}\cr 
}
\tag{3.11}
$$
where the $\L\times\L$-matrix $\RR\left(\cdot\,\rightarrow\cdot\,; C\right)$
has non-zero entries
only for $x,y\in C$ that are given by
$$
\eqalign{
&\RR\left(x\,\rightarrow y\,; C\right)
=\sum_{\srm{\hbox{paths }\g\hbox{ from }x \hbox{ to }y}\atop
{\hbox{ Range}(\g)=C}}\left(\frac{1}{c+2d}\right)^{|\g|+1}\cr
}
\tag{3.12}
$$
For the proof of this formula 
see the Appendix (A.8) and (A.13) ff. 
where also more details
about the random walk expansion can be found. 

Simply from the decomposition of the resolvent
$R_{\L}=R_{G^r}+\left(
R_{\L}-R_{G^r}\right)$ and the random walk representation
for the second term  follows the formula for the centerings
$$
\eqalign{
&m_{\L}^{\tilde m_{\del \L},
\eta_{\L},
\s_{\L}}
=m_{G^r}^{\tilde m_{\del \L},
\eta_{G^r},
\s_{G^r}}
+\sum_{{C \sb \L}\atop
{C\cap(G^r)^c\neq \em}}\bar m(C;
\s_{C},\eta_{C})\cr
}
\tag{3.13}
$$
with
$$
\eqalign{
&m_{G^r}^{\tilde m_{\del \L},
\eta_{G^r},
\s_{G^r}}:=
R_{G^r}\left(
c m^*\s_{G^r}
+\frac{\eta_{G^r}}{q}
+\del_{G^r,\del \L}\tilde m_{\del \L}
\right)\cr
}
\tag{3.14}
$$
and `high-temperature' terms given by the matrix product
$$
\eqalign{
&\bar m(C;
\s_{C},\eta_{C})
=\RR\left(\cdot\,\rightarrow\cdot\,; C\right)
\left(
c m^*\s_{\L}
+\frac{\eta_{\L}}{q}
+\del_{C,\del \L}\tilde m_{\del \L}
\right)\cr
}
\tag{3.15}
$$
From the bound
on the resolvent (A.12) we have uniformly
$$
\eqalign{
&\left|\bar m_{x}(C;
\s_{C},\eta_{C})\right|
\leq \Const (m^*+\d)\left(1+\frac{a}{2d q}  \right)^{-|C|}\cr
}
\tag{3.16}
$$
This quantity is in turn bounded by, say, 
$\left(1+\frac{a}{2d q}  \right)^{-|C|/2}$
if we have that $|C|\geq r$
with 
$r:=\Const \frac{\log m^*}{
\log\left(1+\frac{a}{2d q}  \right)}$.
So we have 
$r\sim \Const \frac{\log m^*}{
\log\left(\frac{1}{q}  \right)}$ for small $q$.

To write both type of summations over connected sets $C$ 
in the same form we note that 
$$
\eqalign{
&\sum_{{C_1 \sb \L\ba\del G}\atop
{C_1\cap(G^r)^c\neq \em,{C_1\cap G^2}\neq \em }}
\del_{\del{G},\L\ba\del{G}}\RR\left(\cdot\,\rightarrow\cdot\,; C_1\right)
\del_{\L\ba\del{G},\del{G}}\cr
&=\sum_{{C_2 \sb \L}\atop
{C_2\cap\del G\neq\em;    C_2\cap(G^r)^c\neq \em}}
\del_{\del{G},\L\ba\del{G}}\RR\left(\cdot\,\rightarrow\cdot\,; C_2\ba \del G
\right)
\del_{\L\ba\del{G},\del{G}}1_{C_2\ba\del G\hbox{ \srm conn.}}
\cr
}
\tag{3.17}
$$
which gives us the same range of summation for both
sort of terms. The expansion then produces triple sums
over connected sets $C$.
Collecting terms according to the union of the occuring $C$'s
we obtain the desired decomposition with
$$
\eqalign{
&\bar H^{\hbox{\srm HT}}_{\del G,G^r }(m_{\del G},
\s_{G^r},\eta_{G^r}
;C, \s_{C},\eta_{C})
\cr
&=-\frac{q}{2}
<\left(m_{\del G}-\bar m_{\del G}^{\s_{G^r}}\right),
\del_{\del{G},\L\ba\del{G}}\RR\left(\cdot\,\rightarrow\cdot\,; C\ba \del G
\right)
\del_{\L\ba\del{G},\del{G}}1_{C\ba\del G\hbox{ \srm conn.}}
\left(m_{\del G}-\bar m_{\del G}^{\s_{G^r}}\right)>
\cr
&+q
<\left(m_{\del G}-\bar m_{\del G}^{\s_{G^r}}\right),
\left(\Pi_{\del G}R_{G^r}\Pi_{\del G}\right)^{-1}
\bar m(C;
\s_{C},\eta_{C})
>\cr
&-q
\sum_{{C_1,C_2 \sb \L; C_1\cup C_2=C}\atop
{C_i\cap\del G\neq\em;    C_i\cap(G^r)^c\neq \em}}\cr
&\quad\times<\left(m_{\del G}-\bar m_{\del G}^{\s_{G^r}}\right),
\del_{\del{G},\L\ba\del{G}}\RR\left(\cdot\,\rightarrow\cdot\,; C_1\ba \del G
\right)
\del_{\L\ba\del{G},\del{G}}1_{C_1\ba\del G\hbox{ \srm conn.}}
\bar m(C_2;
\s_{C_2},\eta_{C_2})>\cr
&+\frac{q}{2}
\sum_{{C_2,C_3 \sb \L; C_2\cup C_3=C}\atop
{C_i\cap\del G\neq\em;    C_i\cap(G^r)^c\neq \em}}
<\bar m(C_2;
\s_{C_2},\eta_{C_2}),
\left(\Pi_{\del G}R_{G^r}\Pi_{\del G}\right)^{-1}
\bar m(C_3;
\s_{C_3},\eta_{C_3})>
\cr
&-\frac{q}{2}
\sum_{{C_1,C_2,C_3 \sb \L; C_1\cup C_2\cup C_3=C}\atop
{C_i\cap\del G\neq\em;    C_i\cap(G^r)^c\neq \em}}\cr
&\quad\times<\bar m(C_2;
\s_{C_2},\eta_{C_2}),
\del_{\del{G},\L\ba\del{G}}
\RR\left(\cdot\,\rightarrow\cdot\,; C_1\ba \del G\right)
\del_{\L\ba\del{G},\del{G}}1_{C_1\ba\del G\hbox{ \srm conn.}}
\bar m(C_3;
\s_{C_3},\eta_{C_3})>
\cr
}
\tag{3.18}
$$
with the short notation $\bar m_{\del G}^{\s_{G^r}}
= m_{G^r}^{\tilde m_{\del \L},
\eta_{G^r},
\s_{G^r}}\Bigl |_{\del G}$.
The bounds are clear
now from the bounds on the resolvent, the choice
of $r$ and the (trivial) control of the $C_i$-sums, i.e.
provided by
$$
\eqalign{
&\sum_{{\hbox{\srm all subsets }S_1,S_2,S_3\sb C}\atop
{\cup_i S_i=C}}
e^{-\a\left(|S_1|+|S_2|+|S_3| \right)}
=\left(3e^{-\a}+3e^{-2\a}+ e^{-3\a}\right)^{|C|}
\leq e^{-\const \a|C|}
\cr
}
\tag{3.19}
$$
\endproof

To proceed with the 
proof of Proposition 3.1 and 
high temperature-expand the $\bar H^{\hbox{\srm HT}}$-terms 
we use the subtraction of bounds-trick 
to ensure the positivity of the resulting activities. 
We thus write for fixed $G$
$$
\eqalign{
&e^{-
\sum_{{C \sb \L}\atop
{C\cap\del G\neq\em;    C \cap(G^r)^c\neq \em}} 
\bar H^{\hbox{\srm HT}}_{\del G,G^r }(m_{\del G},
\s_{G^r},\eta_{G^r}
;C, \s_{C},\eta_{C})
}\cr
&=\prod_{(G^r)_i\hbox{\srm conn. cp. of }G^r}
e^{-\sum_{{C \sb \L; C\hbox{\srm conn. to }(G^r)_i }\atop
{C\cap\del G\neq\em;    C \cap(G^r)^c\neq \em}} 
e^{-\tilde \a|C|}}\cr
&\times e^{
\sum_{{C \sb \L}\atop
{C\cap\del G\neq\em;    C \cap(G^r)^c\neq \em}} 
\left(
n\left(G^r,C \right)
e^{-\tilde \a|C|} -\bar H^{\hbox{\srm HT}}_{\del G,G^r }(m_{\del G},
\s_{G^r},\eta_{G^r}
;C, \s_{C},\eta_{C})
\right)
}
}
\tag{3.20}
$$
where $n(G^r,C)$ is the  number of connected
components of $G^r$ that 
are connected to $C$ (i.e. have $(G^r)_i\cap C\neq \em$).
The exponential in the last line can then be cluster-expanded
and gives 
$$
\eqalign{
&e^{
\sum_{{C \sb \L}\atop
{C\cap\del G\neq\em;    C \cap(G^r)^c\neq \em}} 
\left(
n\left(G^r,C \right)
e^{-\tilde \a|C|} -\bar H^{\hbox{\srm HT}}_{\del G,G^r }(m_{\del G},
\s_{G^r},\eta_{G^r}
;C, \s_{C},\eta_{C})
\right)
}\cr
&=\sum_{{K \sb \L; K=\em\hbox{ or}}\atop
{K\cap\del G\neq\em,    K \cap(G^r)^c\neq \em}} 
\r^{\hbox{\srm HT}}_{\del G,G^r }\left(m_{\del G},
\s_{G^r},\eta_{G^r}
;K, \s_{K},\eta_{K}\right)\cr
}
\tag{3.21}
$$
with 
$0\leq \r^{\hbox{\srm HT}}_{\del G,G^r }\left(m_{\del G},
\s_{G^r},\eta_{G^r}
;K, \s_{K},\eta_{K}\right)
\leq e^{-\tilde\a|K|}$. $\phantom{KKK}$
Here we use the convention that 
$\r^{\hbox{\srm HT}}_{\del G,G^r }\left(m_{\del G},
\s_{G^r},\eta_{G^r}
;K=\em, \s_{K},\eta_{K}\right)=1$.

Note that the resulting activities factorize over 
connected components of
$K\cup G^r$;
this is due to the (trivial) fact that the number $n(G^r,C)$
that enters the definition of the contour activities
depends only on those components of $G^r$ that $C$
is connected to. 
We put
$$
\eqalign{
&\rho^{\hbox{\srm geo}}\left(\del G,G^r\right)
:=\prod_{(G^r)_i\hbox{\srm conn. cp. of }G^r}
e^{-\sum_{{C \sb \L; C\hbox{\srm conn. to }(G^r)_i }\atop
{C\cap\del G\neq\em;    C \cap(G^r)^c\neq \em}} 
e^{-\tilde \a|C|}}\cr
}
\tag{3.22}
$$
and note that
$$
\eqalign{
&1\geq \rho^{\hbox{\srm geo}}\left(\del G,G^r\right)
\geq e^{-|G^r|e^{-\const \tilde \a } }\cr
}
\tag{3.23}
$$
We can finally carry out the integral on $\del G$
to get the form as promised in the proposition.
In doing so it is convenient to
pull out a normalization constant
and introduce the normalized Gaussian measures on $\del G$
corresponding to the Hamiltonian on the r.h.s. 
of (3.8), given by 
$$
\eqalign{
&\int\mu^{\left(\tilde m_{\del_{\del \L} G^r},
0_{\del_{\L} G^r}\right),
\eta_{G^r},\s_{G^r}}_{\del G,G^r}(dm_{\del G})f(m_{\del G})\cr
&:=\frac{
\int dm_{\del G}
e^{-\D H^{\left(\tilde m_{\del_{\del \L} G^r},
0_{\del_{\L} G^r}\right),
\eta_{G^r},\s_{G^r}}_{\del G,G^r}\left(m_{\del G} \right)
}f(m_{\del G}) }
{\int dm'_{\del G}
e^{-\D H^{\left(\tilde m_{\del_{\del \L} G^r},
0_{\del_{\L} G^r}\right),
\eta_{G^r},\s_{G^r}}_{\del G,G^r}\left(m'_{\del G} \right)}}
\cr
}
\tag{3.24}
$$
So we can write
$$
\eqalign{
&\int dm_{\del G}
e^{-\D H^{\tilde m_{\del \L},
\eta_{\L},\s_{\L}}_{\del G,\L}\left(m_{\del G} \right)}
1_{m_{\del G}\in U^{\del G}}
I^{\tilde m_{\del \L},m_{\del G},
\eta_{G},\s_{G}}_{G}\cr
&=\left(2 \pi\right)^{\frac{|\del G|}{2}}
\sqrt{\det \left(\Pi_{\del G}\left(a-q\D_{G^r} \right)^{-1}\Pi_{\del G}\right)}
\sum_{{K \sb \L; K=\em\hbox{ or}}\atop
{K\cap\del G\neq\em,    K \cap(G^r)^c\neq \em}}
\rho^{\hbox{\srm geo}}\left(\del G,G^r\right)
\cr
&\int\mu^{\left(\tilde m_{\del_{\del \L} G^r},
0_{\del_{\L} G^r}\right),
\eta_{G^r},\s_{G^r}}_{\del G,G^r}(dm_{\del G})\,\,
1_{m_{\del G}\in U^{\del G}}\,\,
\r^{\hbox{\srm HT}}_{\del G,G^r }\left(m_{\del G},
\s_{G^r},\eta_{G^r}
;K, \s_{K},\eta_{K}\right)
I^{\tilde m_{\del \L},m_{\del G},
\eta_{G},\s_{G}}_{G}\cr
}
\tag{3.25}
$$
This has in fact the desired form (3.5)
with the obvious definition
$$
\eqalign{
&\r^{\tilde m_{\del_{\del \L} \tilde G}}\left(G,\tilde G; \s_{\tilde G},\eta_{\tilde G} \right)
:=\rho^{\hbox{\srm geo}}\left(\del G,G^r\right)\cr
&\times\int\mu^{\left(\tilde m_{\del_{\del \L} G^r},
0_{\del_{\L} G^r}\right),
\eta_{G^r},\s_{G^r}}_{\del G,G^r}(dm_{\del G})\,\,
1_{m_{\del G}\in U^{\del G}}\,\,
\r^{\hbox{\srm HT}}_{\del G,G^r }\left(m_{\del G},
\s_{G^r},\eta_{G^r}
;K, \s_{K},\eta_{K}\right)
I^{\tilde m_{\del \L},m_{\del G},
\eta_{G},\s_{G}}_{G}\cr
}
\tag{3.26}
$$
with $K=\tilde G\ba G^r$ on the r.h.s.
Note that these activities factorize over
connected components of $\tilde G$.

In view of the trivial bound (3.23) 
on the geometric activity (3.22)
and the normalization of the measure,
the bounds follows from the HT-bounds and
the bounds on the anharmonic activities $I$.
The value of the `Peierls constant' $\bar\a$  
is now clear from $\bar \a= \Const \min\{
(2r+1)^{-d}\log\frac{1}{\e},\tilde\a \}$, assuming that 
both terms in the minimum are sufficiently 
large.\endproof

To finish with the proof of Proposition 3.1
is now an easy matter. 
Using the formula for the determinant from Appendix  
(A.3) we can write 
$$
\eqalign{
&\frac{1}{\det \left(a-q \D_{\L\ba\ov G} \right)}
\det \left(\Pi_{\del G}\left(a-q\D_{G^r} \right)^{-1}\Pi_{\del G}\right)\cr
&=\frac{1}{\det \left(a-q \D_{\L} \right)}
\times
\frac{\det \left(\Pi_{\del G}\left(a-q\D_{G^r} \right)^{-1}
\Pi_{\del G}\right)}
{\det \left(\Pi_{\del G}\left(a-q\D_{\L} \right)^{-1}
\Pi_{\del G}\right)}
\times 
\det \left(a-q \D_{G} \right)\cr
}
\tag{3.27}
$$
Remember that the 
correction given by the middle term on the r.h.s. stems 
from the lack
of terms with range longer than $r$ in the quadratic form
of (3.24) that we had cut off. 
The random walk representation then gives 
the following expansion whose proof is given in the Appendix.

\lemma{3.4}{
$$
\eqalign{
\frac{\det \left(\Pi_{\del G}\left(a-q\D_{G^r} \right)^{-1}
\Pi_{\del G}\right)}
{\det \left(\Pi_{\del G}\left(a-q\D_{\L} \right)^{-1}
\Pi_{\del G}\right)}
=e^{-2\sum_{{C \sb \L}\atop
{C\cap\del G\neq\em;    C \cap(G^r)^c\neq \em}}
\e^{\hbox{\srm det}}(C)}
}
\tag{3.28}
$$
where
$0\leq \e^{\hbox{\srm det}}(C)\leq e^{-\a|C|}$
with $\a\sim \const\log\frac{1}{q}$.
} 

Next we use
subtraction of bounds as in (3.20) to write 
$$
\eqalign{
&e^{-\sum_{{C \sb \L}\atop
{C\cap\del G\neq\em;    C \cap(G^r)^c\neq \em}}
\e^{\hbox{\srm det}}(C)}\cr
&=\rho^{\hbox{\srm geo,det}}
\left(\del G,G^r\right)\sum_{{K \sb \L; K=\em\hbox{ or}}\atop
{K\cap\del G\neq\em,    K \cap(G^r)^c\neq \em}} 
\r^{\hbox{\srm det}}_{\del G,G^r }\left(K\right)\cr
}
\tag{3.29}
$$
where 
$1\geq \rho^{\hbox{\srm geo,det}}
\geq e^{-|G^r|e^{-\const \tilde \a } }$
and 
$0\leq \r^{\hbox{\srm det}}_{\del G,G^r }\left(K\right)
\leq e^{-\const\tilde \a|K|}$.
So we get 
$$
\eqalign{
&
e^{b|\L|}Z_{\L}^{\tilde m_{\del \L}}(\s_{\L})
=\left(2\pi \right)^{\frac{|\L|}{2}}
\left(\det \left(a-q \D_{\L} \right)\right)^{-\frac{1}{2}}
e^{-\inf_{m'_{\L}} H^{\tilde m_{\del \L},\eta_{\L}, 
\s_\L}_{\L}\left(m'_{\L}\right)}\cr
&\times\sum_{G:\em\sb G \subset \L } 
\left(2\pi \right)^{-\frac{|G|}{2}}\sqrt{\det \left(a-q \D_{G} \right)}
\sum_{{\tilde G:\tilde G\sb \L}\atop{G^r\sb \tilde G}}
\r^{\tilde m_{\del_{\del \L} \tilde G}}\left(G,\tilde G; \s_{\tilde G},\eta_{\tilde G} \right)\cr
&\times\rho^{\hbox{\srm geo,det}}
\left(\del G,G^r\right)\sum_{{K \sb \L; K=\em\hbox{ or}}\atop
{K\cap\del G\neq\em,    K \cap(G^r)^c\neq \em}} 
\r^{\hbox{\srm det}}_{\del G,G^r }\left(K\right)\cr
}
\tag{3.30}
$$
This can be summed over $G,\tilde G,K$
(collecting terms that give the same $\tilde G\cup K$)
to yield the claims of Proposition 3.1.
\endproof


\overfullrule=0pt
\bigskip\bigskip
\chap{IV. The effective contour model: Gaussian case}

It is instructive to make explicit the result of 
our transformation to an effective Ising-contour model
at first without the presence of anharmonic potentials
where the proof is easy. 
In fact, as we will explain in  Chapter V, 
the work done in Chapters II and III will then imply
that a weak anharmonicity 
can be absorbed in essentially 
the same type of contour activities 
we encounter already in the purely Gaussian model.
 
We remind the reader that in the purely Gaussian
case the Ising-weights 
$\left(T\left(\mu_{\L}^{\tilde m_{\del \L},\eta_{\L}}\right)\right)
\left(\s_{\L}\right)$ are
obtained by normalizing  
$\exp\left(-\inf_{m_{\L}\in \R^{\L}} H^{\tilde m_{\del \L},\eta_{\L}, 
\s_\L}_{\L}\left(m_{\L}\right)\right)$ by its $\s_{\L}$-sum. 
For simplicity we restrict now to the boundary condition
$\tilde m_x=m^*$ for all $x$
(that is everywhere in the minimum of the positive wells).

We will now express the latter exponential
as a sum over contour-weights.
To do so we use the following (by now standard)
definition of a signed contour model, including 
$+$-boundary conditions.

{\bf Definition: }{\it A {\rm contour} in $\L$ 
is a pair $\G=\left(\un\G,\s_{\L} \right)$
where $\un\G\sb \L$ (the support of $\G$)
and  the spin-configuration $\s_{\L}\in \{-1,1\}^{\L}$
are such that the extended 
configuration $(\s_{\L},+1_{\Z^d\ba \L})$ is 
constant on connected components of $\Z^d\ba \un\G$.

The {\rm connected components} of 
a contour $\G$ are the contours $\G_i$ whose 
supports are the connected components ${\un \G}_i$ of $\un \G$
and whose sign is determined by 
the requirement that it be the same as that of $\G$ on $\ov{{\un\G}_i}$.

A {\rm contour model representation} for a probability measure 
$\nu$ on the space $\{-1,1\}^{\L}$ of Ising-
spins in $\L$ 
is a probability measure $N$ 
on the space of 
contours in $\L$ s.t. the marginal on the spin reproduces
$\nu$, i.e. we have 
$$
\eqalign{
&\nu(\{\s_{\L}\})= \sum_{{\G}\atop {\s_{\L}\left(\G\right)=\s_{\L}}}
N\left(\{\G\}\right)
}
\tag{4.1}
$$ 
}

Recall that, in the simplest low-temperature 
contour model, arising from 
the standard nearest neighbor ferromagnetic Ising model,
$N\left(\{\G\}\right)=\Const\times\r(\G)$ is proportional to 
a (non-negative) activity $\r(\G)$ that factorizes over connected components
of the contour and obeys a Peierls estimate of the form
$\r(\G)\leq e^{-\t|\un\G|}$. 
There is a satisfying theory for the 
treatment of deterministic models 
with additional 
volume terms for activities that are not 
necessarily symmetric under spin-flip, known as
Pirogov-Sinai theory.
For random models then, while the activities will be random, 
there have to be also  
additional random volume-contributions
to $N\left(\{\G\}\right)$, even 
when the distribution of the disorder is symmetric, 
caused by local fluctuations in the free energies 
of the different states. 
The fluctuations of these volume terms are responsible for 
the fact that, even in situations where the disorder
is `irrelevant', not all 
contours carry exponentially small mass but the formation of 
some contours (depending on the specific realization) is favorable. 
It is the control of this phenomenon that 
poses the difficulties in the analysis of the stability
of disordered contour models and necessitates RG
(or possibly some related multiscale method).

To write down the Peierls-type estimates to come  
for the present model we 
introduce the `naive contour-energy'  (i.e.
the $d-1$-dimensional volume of the plaquettes
separating plus- and minus-regions in $\Z^d$) putting
$$
\eqalign{
E_s(\G)=\sum_{\{x,y\}\sb \un \G, d(x,y)=1}
1_{\s_{x}\neq \s_{y}}
+\sum_{{x\in \un \G,y\in \del \L}\atop {d(x,y)=1}}
1_{\s_{x}=-1}
\cr
}
\tag{4.2}
$$
again taking into the interaction
with the positive boundary condition.

Then the result of the transformation of the
purely Gaussian continuous  spin model
to an effective Ising-contour model is given 
by the following

\proposition{4.1}{\it  
Suppose that $q$ is sufficiently small,
$q(m^*)^2$ sufficiently large, $a$ is of the order $1$
and $\d\leq \Const m^*$ with a constant of the order $1$. 

Then there  is a $\s_{\L}$-independent constant 
$K_{\L}\left(\eta_{\L}\right)$
s.t. we have  the representation 
$$
\eqalign{
&e^{-\inf_{m_{\L}\in \R^{\L}} H^{+m^* 1_{\del \L},\eta_{\L}, 
\s_\L}_{\L}\left(m_{\L}\right)}
=K_{\L}\left(\eta_{\L}\right)\cr
&\times e^{\sum_{C\sb V^+(\s_{\L})}S^{\hbox{\srm Gau\ss}}_{C}(\eta_C)
-\sum_{C\sb V^-(\s_{\L})}S^{\hbox{\srm Gau\ss}}_{C}(\eta_C)
}
\sum_{{\G}\atop {\s_{\L}\left(\G\right)=\s_{\L}}}
\r_0(\G; \eta_{\un\G})
}
\tag{4.3}
$$
for any $\s_{\L}$, with $V^\pm(\s_{\L})=\{x\in \L; \s_x=\pm 1\}$. 
Here

\item{(i)} $\eta_C\mapsto S^{\hbox{\srm Gau\ss}}_{C}(\eta_C)$ 
are functions of the random fields indexed 
by the connected sets $C\sb \L$.
They are symmetric, i.e.
$S^{\hbox{\srm Gau\ss}}_C(-\eta_C)=-S^{\hbox{\srm Gau\ss}}_C(\eta_C)$
and invariant under lattice-shifts.
For $C=\{x\}$ we have in particular $S^{\hbox{\srm Gau\ss}}_x(\eta_x)=\frac{a m^*}{a+2dq }\eta_x$.

\item{(ii)}The activity $\r_{0}(\G;\eta_{\un\G})$ 
is non-negative.
It factorizes over the connected components of $\G$, i.e.
$$
\eqalign{
&\r_{0}(\G;\eta_{\un \G})=
\prod_{\G_i\hbox{ conn cp. of }\G}
\r_{0}(\G_i;\eta_{\G_i})
\cr
}
\tag{4.4}
$$

\noindent For $\un\G$ not touching the boundary (i.e. 
$\del_{\del \L}\un\G=\em$) the value of 
$\r_{0}(\G;\eta_{\un\G})$ is independent of $\L$.
We then have the `infinite volume
properties' of

\item{(a)} Spin-flip symmetry, i.e.  
$\r_{0}((\un\G,\s_{\L});\eta_{\un \G})
=\r_{0}((\un\G,-\s_{\L});-\eta_{\un \G})$

\item{(b)} Invariance under joint lattice shifts
of spins and random fields  

\noindent {\bf Peierls-type bounds:}
There exist positive constants $\tilde \b_{\hbox{\srm Gau\ss}},\b$ s.t. 
we have the bounds
$$
\eqalign{
&0\leq \r_{0}(\G;\eta_{\un \G})
\leq
e^{-\b E^s(\G)-\tilde \b_{\hbox{\srm Gau\ss}} |\un \G|  }
\cr
}
\tag{4.5}
$$
uniformly in $\eta_{\un \G}\in [-\d,\d]^{\un \G}$
where the `Peierls-constants' can be chosen like
$$
\eqalign{
&\b=\frac{q (m^*)^2}{2} \frac{a^2}{(a+2dq)^2-q^2},\quad
\tilde \b_{\hbox{\srm Gau\ss}}=\Const \times \min\left\{
\log\frac{1}{q}, q{m^*}^2 \left(\frac{\log \frac{1}{q}}{\log m^*}
\right)^d\right\}-m^*\d
\cr
}
\tag{4.6}
$$

\noindent The non-local random fields obey the estimate
$$
\eqalign{
&\left|S^{\hbox{\srm Gau\ss}}_C(\eta_C)\right|\leq  \d m^* e^{-\tilde \b_0|C|}
\cr
}
\tag{4.7}
$$
for all $|C|\geq 1$ with $\tilde \b_0=\Const \times\log\frac{1}{q}$.
}

\noindent {\bf Remark 1: }This structure
will be familiar to the reader familiar with [BK1] or 
[BoK] (see page 457).
Indeed, the above model falls in the class of 
contour models given in (5.1) of [BK1] 
(as written therein for the partition
function). This form was then shown to be of 
sufficient generality to describe the contour models arising from 
the random field Ising model under any iteration
of the contour-RG that was constructed in [BK1]. 
(The additional  non-local interaction $W(\G)$
encountered in [BK1] is not necessary and could be expanded by 
subtraction-of-bounds as in (3.20), giving rise
to enlarged supports $\G$, as it was done in [BoK]).

\noindent {\bf Remark 2: }There 
is some freedom in the precise formulation of contours and
contour activities, resp. the  question 
of keeping information
additional to the support and the spins 
on the contours. 
[BK1] speak of inner and outer supports, while
in [BoK] it was preferred to define contours
with activities containing interactions. 
The latter is motivated by the limit of the temperature
going to zero (making the interactions vanish). 
Since we do not perform such
a limit here, we present the simplest 
possible choice and do not make such distinctions here,
simply collecting all interactions from 
different sources into `the support'.

\noindent {\bf Remark 3: }
The magnitude of $\b\sim \Const q {m^*}^2$ is easily understood
since it gives the true order of magnitude
of the minimal energetic contribution
to the original Hamiltonian of a nearest neighbor pair
of continuous spins sitting in potential wells
with opposite signs. 
This term appears again in the estimate on 
$\tilde \b_{\hbox{\srm Gau\ss}}$ (up to logarithmic
corrections) together with 
a contribution of the same form as $\tilde \b_0$.
The latter comes from a straight-forward expansion
of long-range contributions. 
The last term in (4.6),  $m^*\d$, is a trivial control 
on the worst realization of the random fields;
it could easily be avoided by the introduction 
of so-called `bad regions'.
These are regions of space where the 
realizations of the random fields are
exceptionally (and dangerously) large in some sense
and, while comparing with [BK1] or [BoK],
the reader might have already missed them. 
Indeed, a renormalization of the present model
will immediately produce such bad regions in the next steps.
Of course, we could have started, here and also 
in the presence of anharmonicity, 
with an unbounded distribution of the $\eta_x$.
In the latter case we would have to single out regions of space
where the behavior of our transformation to the 
Ising-model gets exceptional (i.e. because 
we lose Lemma 2.5.)
We chose however not to treat this case here
in order to keep the technicalities down.


\proof An elementary computation yields the important 
fact that the minimum of the quadratic
Hamiltonian (2.6) with any boundary condition
$\tilde m$ is given by
$$
\eqalign{
&-\inf_{m_{\L}\in \R^{\L}} H^{\tilde m_{\del \L},\eta_{\L}, 
\s_\L}_{\L}\left(m_{\L}\right)
=-\frac{ a^2 (m^*)^2}{2 q}<\s_\L,R_{\L}\s_\L>_{\L}
+\frac{a(m^*)^2}{2}|\L|\cr
&-\frac{a m^*}{q}<\eta+\tilde\eta_{\del(\L^c)}(q\tilde m),R_\L\s_\L>_{\L}\cr
&\quad
-\frac{1}{2q}<\eta+\tilde\eta_{\del(\L^c)}(q\tilde m), 
R_\L\left(
\eta+\tilde\eta_{\del(\L^c)}(q\tilde m)
\right)>_{\L}
+\frac{q}{2}\sum_{{x\in \L; y\in \del \L}
\atop{d(x,y)=1}}\tilde m_y^2
\cr
}
\tag{4.8}
$$ 
with $\tilde\eta_{\del(\L^c)}(\tilde m):=
\del_{\L,\del\L}\tilde m_{\del\L}$ denoting the field
created by the boundary condition. 
We subtract a term that is constant for $\s_{\L}$
(and thus of no interest) and write 
$$
\eqalign{
&\inf_{m_{\L}} H^{\tilde m_{\del \L},\eta_{\L}, 
\s_\L}_{\L}
\left(m_{\L}\right)
-\inf_{m_{\L}} H^{\tilde m_{\del \L},\eta_{\L}, 
1_\L}
\left(m_{\L}\right)
-\frac{a m^*}{q}<\eta_\L,R_\L 1_\L>_{\L}
\cr
&=-\frac{ a^2 (m^*)^2}{2 q}
\left(<\s_\L,R_{\L}\s_\L>_{\L}
-<1_\L,R_{\L}1_\L>_{\L}\right)
-\frac{a m^*}{q}<\eta_\L,R_\L\s_\L>_{\L}\cr
&-a m^*<\tilde\eta_{\del(\L^c)}(\tilde m),R_\L\left(
\s_\L-1_\L\right)>_{\L}\cr
}
\tag{4.9}
$$
The first term on the r.h.s. gives rise
to the low-temp. Peierls constant;
the next term is a weakly nonlocal 
random field term (suppressed by the decay of
the resolvent) and the last term the 
symmetry-breaking coupling
to the boundary.
As in Chapter III 
we use the random walk representation 
$R_{\L}=\sum_{C\sb \L}
\RR\left(\cdot\,\rightarrow\cdot\,; C\right)$
(see Appendix (A.11))and decompose according
to the size of $C$'s. 

As the first step for the contour representation
we associate to any spin-configuration 
$\s_{\L}\in \{-1,1\}^{\L}$ a preliminary (or `inner') support
in the following way. 
Choose some finite integer $r\geq 1$, 
to be determined below, and put
$$
\eqalign{
&\un{\G}^+_{\L}(\s_\L):=
\{x\in \L;
\exists y\in \L
\hbox{ s.t. }d(x,y)\leq r
\hbox{ where }\s_x\neq \s_y 
\}\cr
&\quad\cup \{x\in \L; d(x,\del \L)\leq r+1
\hbox{ where }\s_x=-1
\}
}
\tag{4.10}
$$
The second term makes this definition 
$\L$-dependent by taking 
into account the interaction with the boundary
leading to the (desired) 
symmetry breaking for contours touching the boundary. 
For given $\s_{\L}$ the activities $\r_0(\G; \eta_{\un\G})$
to be defined will be non-zero only for supports
$\un\G\sp \un{\G}^{+}(\s_\L)$.
The range $r$ will be chosen below in such a way that the 
terms corresponding to interactions with range larger 
than $r$ have decayed sufficiently so that they 
can be high-temperature expanded in a straightforward
way. This choice then also determines the 
value of the Peierls-constant for the low-temperature 
contributions.  

Keeping the small $C$'s of 
diameter up to $r$ define the (preliminary) `low-temperature
activities' 
$$
\eqalign{
&\r^{\hbox{\srm LT},\tilde m_{\del \L}}(\s_{\L})\cr
&:=e^{\sum_{C\sb \L;\diam(C)\leq r }\Bigl[
\frac{ a^2 (m^*)^2}{2 q}\left( 
<\s_{C},\RR\left(\cdot\,\rightarrow\cdot\,; C\right)\s_{C}>
-<1_{C},\RR\left(\cdot\,\rightarrow\cdot\,; C\right)1_{C}>
\right)
+a m^*<\tilde\eta_{\del(\L^c)}(\tilde m),
\RR\left(\cdot\,\rightarrow\cdot\,; C\right)
\left(
\s_{C}-1_{C}\right)>
\Bigr]}\cr
}
\tag{4.11}
$$
Note that the `inner support' (4.10)
can be trivially rewritten as 
$$
\eqalign{
&\un{\G}^{+}(\s_\L)= \bigcup_{{C\sb\L;\diam(C)\leq r}\atop{
\s_{C}\neq 1_{C}\hbox { \srm and }\s_{C}\neq -  1_{C}}
}C
\cup \bigcup_{{C\hbox{ \srm conn. to }\del \L}\atop{
\diam(C)\leq r;\s_{C}\neq 1_{C}}
}C
}
\tag{4.12}
$$
which shows that it 
is just the union of all connected $C$'s with diameter
less or equal $r$ that give 
any contribution to the sum occuring in the exponent
of (4.11).
So we can rewrite 
$$
\eqalign{
&e^{-\inf_{m_{\L}} H^{\tilde m_{\del \L},\eta_{\L}, 
\s_\L}_{\L}
\left(m_{\L}\right)
+\inf_{m_{\L}} H^{\tilde m_{\del \L},\eta_{\L}, 
1_\L}
\left(m_{\L}\right)
+\frac{a m^*}{q}<\eta_\L,R_\L 1_\L>_{\L}}
\cr
&=\r^{\hbox{\srm LT},\tilde m_{\del \L}}(\s_{\L})
e^{\frac{a m^*}{q}\sum_{{C\sb \L;\diam(C)\leq r}\atop
{\s_{C}\neq \const} }<\eta_C,
\RR\left(\cdot\,\rightarrow\cdot\,; C\right)\s_{C}>}\cr
&e^{\frac{a m^*}{q}\sum_{C\sb V^+(\G)}
<\eta_C,
\RR\left(\cdot\,\rightarrow\cdot\,; C\right)1_{C}>
-\frac{a m^*}{q}\sum_{C\sb V^-(\G)}
<\eta_C,
\RR\left(\cdot\,\rightarrow\cdot\,; C\right)1_{C}>
}\cr
&e^{\sum_{{C\sb \L;\diam(C)> r}\atop
{\s_{C}\neq \const}}\Bigr[
\frac{ a^2 (m^*)^2}{2 q}\left( 
<\s_{C},\RR\left(\cdot\,\rightarrow\cdot\,; C\right)\s_{C}>
-<1_{C},\RR\left(\cdot\,\rightarrow\cdot\,; C\right)1_{C}>
\right)
+\frac{a m^*}{q}
<\eta_C,
\RR\left(\cdot\,\rightarrow\cdot\,; C\right)\s_{C}>
\Bigl]
}\cr
&e^{
\sum_{{C\sb \L;\diam(C)> r}\atop{C\cap \del (\L)^c\neq \em} }
a m^*<\tilde\eta_{\del(\L^c)}(\tilde m),
\RR\left(\cdot\,\rightarrow\cdot\,; C\right)
\left(\s_{C}-1_{C}\right)>
}\cr
}
\tag{4.13}
$$
The terms in the first line 
depend only on quantities on $\un{\G}^{+}(\s_\L)$ 
and factorize over its connected components.
They will give contributions  
to the activities $\r_{0}$.
The terms in the second line are the small-field
contributions to the vacua given by 
$$
\eqalign{
&S^{\hbox{\srm Gau\ss}}_{C}(\eta_C):=\frac{a m^*}{q}<\eta_C,
\RR\left(\cdot\,\rightarrow\cdot\,; C\right)1_{C}>
}
\tag{4.14}
$$
The terms in the last two lines 
are small (since only $C$'s with sufficiently large
diameter contribute) 
and only non-zero
for $C$'s intersecting with $\un{\G}^{+}(\s_\L)$
or touching the boundary.
They can be expanded.

Let us see now what explicit bounds we get on the low-temperature
activity (4.11). 
Keeping only $C$'s made of two nearest neighbors
$x,y=x+e$ we have the upper bound
$$
\eqalign{
&\sum_{C\sb \L;\diam(C)\leq r }\frac{ a^2 (m^*)^2}{2 q}\left( 
<\s_{C},\RR\left(\cdot\,\rightarrow\cdot\,; C\right)\s_{C}>
-<1_{C},\RR\left(\cdot\,\rightarrow\cdot\,; C\right)1_{C}>
\right)\cr
&\leq -\frac{ a^2 (m^*)^2}{q}\sum_{\{x,y\}\sb \un{\G}^{+}(\s_\L), d(x,y)=1}
\RR\left(x\,\rightarrow\,y; C=\{x,y\}\right)1_{\s_{x}\neq \s_{y}}
\cr
}
\tag{4.15}
$$
Computing 
$$
\eqalign{
&\RR\left(x\,\rightarrow\,x+e; C=\{x,x+e\}\right)
=\frac{1}{c+2d}\sum_{k=1,3,5,\dots}^\infty \left(
\frac{1}{c+2d}\right)^k
=\frac{1}{(c+2d)^2-1} 
\cr
}
\tag{4.16}
$$
with $c=a/q$ we get an upper bound on the l.h.s. of (4.15) of 
$-2\b
\sum_{\{x,y\}\sb \un{\G}^{+}(\s_\L), d(x,y)=1}1_{\s_{x}\neq \s_{y}}
$
where $\b$ is given by (4.6).
Applying a similar reasoning on the boundary term,
thereby using that $\RR\left(x\,\rightarrow\,; C=\{x\}\right)
=\frac{1}{c+2d}$, gives the bound
$$
\eqalign{
&\sum_{C\sb \L;\diam(C)\leq r }
a m^*<\tilde\eta_{\del(\L^c)}(\tilde m),
\RR\left(\cdot\,\rightarrow\cdot\,; C\right)
\left(\s_{C}-1_{C}\right)>\cr
&\leq 
-q (m^*)^2 \frac{2 a}{a+2dq}
\sum_{{x\in \un{\G}^{+}(\s_\L),y\in \del \L}\atop {d(x,y)=1}}
1_{\s_{x}=-1}
}
\tag{4.17}
$$
Since the modulus of the
prefactor in the last line is larger than $2\b$ we get 
an energetic suppression of 
$$
\eqalign{
&\r^{\hbox{\srm LT},\tilde m_{\del \L}}(\s_{\L})
\leq e^{- 2\b E_s(\G^+_{\L}(\s_{\L}),\s_{\L})}
\leq e^{- \b E_s(\G^+_{\L}(\s_{\L}),\s_{\L}) 
-\b (2r+1)^{-d}|\G^+_{\L}(\s_{\L})|
}\cr
}
\tag{4.18}
$$
Using $\sum{y}R_{\L;x,y}\leq 1/c$ for the next
term in (4.13) we have immediately
$$
\eqalign{
&\frac{a m^*}{q}\sum_{{C\sb \L;\diam(C)\leq r}\atop
{\s_{C}\neq \const} }<\eta_C,
\RR\left(\cdot\,\rightarrow\cdot\,; C\right)\s_{C}>
\leq m^* \d|\G^+_{\L}(\s_{\L})|\cr
}
\tag{4.19}
$$
This finishes the Peierls estimate for the low-temperature
contributions.

Let us come to the treatment of the `high-temperature
parts' in (4.13) now, proceeding algebraically at first. 
Using subtraction-of-bounds 
as in Chapter III (3.20) we get the high-temperature
expansion
$$
\eqalign{
&e^{\sum_{{C\sb \L;\diam(C)> r}\atop
{\s_{C}\neq \const}}\Bigr[
\frac{ a^2 (m^*)^2}{2 q}\left( 
<\s_{C},\RR\left(\cdot\,\rightarrow\cdot\,; C\right)\s_{C}>
-<1_{C},\RR\left(\cdot\,\rightarrow\cdot\,; C\right)1_{C}>
\right)
+\frac{a m^*}{q}
<\eta_C,
\RR\left(\cdot\,\rightarrow\cdot\,; C\right)\s_{C}>
\Bigl]
}\cr
&=\tilde\rho^{\hbox{\srm geo}}\left(\un{\G}^{+}(\s_\L)\right)
\sum_{{K\sb \L;\diam(K)> r \hbox{ or }K=\em}\atop
{\s_{K}\neq \const}}
\r^{\hbox{\srm HT1}}\left(K, \s_{K},\eta_{K}\right)
}
\tag{4.20}
$$
if the terms in the exponential on the l.h.s.
are sufficiently small. 
To control them we just use the bound (A.12)
$$
\eqalign{
&\sum_{y\in \Z^d}\RR\left(x\,\rightarrow y\,; C\right)
\leq \frac{1}{c}\left(\frac{2d}{c+2d}\right)^{|C|-1}\cr
}
\tag{4.21}
$$
This gives the deterministic bound 
upper bound on the first two terms in (4.13)
of 
$$
\eqalign{
&\frac{ a^2 (m^*)^2}{2 q}\left| 
<\s_{C},\RR\left(\cdot\,\rightarrow\cdot\,; C\right)\s_{C}>
-<1_{C},\RR\left(\cdot\,\rightarrow\cdot\,; C\right)1_{C}>
\right|\leq e^{-\a|C|}
}
\tag{4.22}
$$
if we have
$$
\eqalign{
&\a\leq \log\left(1+\frac{a}{2dq}  \right)
\left[1-\frac{\log(a{m^*}^2)}
{(|C|-1)\log\left(1+\frac{a}{2dq}  \right)}\right]
-\frac{1}{e}
}
\tag{4.23}
$$
which is in turn bounded by $\a_0:=\frac{1}{2}
\log\left(1+\frac{a}{2dq}\right)
-\frac{1}{e}$ for the $C$'s in the above sum if we put 
$$
\eqalign{
&r=\left[
2\frac{\log(a{m^*}^2)}
{\log\left(1+\frac{a}{2dq}  \right)}
\right]+1\sim 4 \frac{\log m^*}{\log\frac{1}{q}}
}
\tag{4.24}
$$
Remember here that we are interested in the regime 
of $\frac{1}{q}$ small and ${m^*}^2$ even larger. 

Assuming (4.24), with $\left|\eta_x\right|\leq \d$ 
the random field contribution is estimated by
$$
\eqalign{
&\frac{a m^*}{q}\left|<\eta_C,
\RR\left(\cdot\,\rightarrow\cdot\,; C\right)\s_{C}>\right|\leq 
\frac{\d}{a m^*}e^{-\a_0|C|}
}
\tag{4.25}
$$
where can use that $\frac{\d}{a m^*}\leq \Const$.
The estimates on $S^{\hbox{\srm Gau\ss}}_C(\eta_C)$ are obtained 
in the very same way.

In passing we verify that all activities 
constructed so far are invariant under 
joint flips of spins and random fields (inside $\L$). 
The boundary terms can be expanded similarly giving 
$$
\eqalign{
&e^{
\sum_{{C\sb \L;\diam(C)> r}\atop{C\cap \del (\L)^c\neq \em} }
a m^*<\tilde\eta_{\del(\L^c)}(\tilde m),
\RR\left(\cdot\,\rightarrow\cdot\,; C\right)
\left(\s_{C}-1_{C}\right)>
}\cr
&=e^{
- 2\sum_{{C\sb \L;\diam(C)> r}\atop{C\cap \del (\L)^c\neq \em} }
a m^*<\tilde\eta_{\del(\L^c)}(\tilde m),
\RR\left(\cdot\,\rightarrow\cdot\,; C\right)1_{C}>}
e^{
\sum_{{C\sb \L;\diam(C)> r}\atop{C\cap \del (\L)^c\neq \em} }
a m^*<\tilde\eta_{\del(\L^c)}(\tilde m),
\RR\left(\cdot\,\rightarrow\cdot\,; C\right)
\left(\s_{C}+1_{C}\right)>}\cr
&=e^{
- 2\sum_{{C\sb \L;\diam(C)> r}\atop{C\cap \del (\L)^c\neq \em} }
a m^*<\tilde\eta_{\del(\L^c)}(\tilde m),
\RR\left(\cdot\,\rightarrow\cdot\,; C\right)1_{C}>}\cr
&\quad\times
\sum_{{K\sb \L;\diam(K)> r\hbox{ or }K=\em}
\atop{K\cap \del (\L)^c\neq \em} }
\r^{\hbox{\srm HT2}}\left(K, \s_{K},\eta_{K}\right)
}
\tag{4.26}
$$
This gives
$$
\eqalign{
&e^{-\inf_{m_{\L}} H^{\tilde m_{\del \L},\eta_{\L}, 
\s_\L}_{\L}
\left(m_{\L}\right)
+\inf_{m_{\L}} H^{\tilde m_{\del \L},\eta_{\L}, 
1_\L}_{\L}
\left(m_{\L}\right)
+\frac{a m^*}{q}<\eta_\L,R_\L 1_\L>_{\L}
+2\sum_{{C\sb \L;\diam(C)> r}\atop{C\cap \del (\L^c)\neq \em} }
a m^*<\tilde\eta_{\del(\L^c)}(\tilde m),
\RR\left(\cdot\,\rightarrow\cdot\,; C\right)1_{C}>}
\cr
&=\r^{\hbox{\srm LT},\tilde m_{\del \L}}(\s_{\L})
e^{\frac{a m^*}{q}\sum_{{C\sb \L;\diam(C)\leq r}\atop
{\s_{C}\neq \const} }<\eta_C,
\RR\left(\cdot\,\rightarrow\cdot\,; C\right)\s_{C}>}\cr
&e^{\frac{a m^*}{q}\sum_{C\sb V^+(\G)}
<\eta_C,
\RR\left(\cdot\,\rightarrow\cdot\,; C\right)1_{C}>
-\frac{a m^*}{q}\sum_{C\sb V^-(\G)}
<\eta_C,
\RR\left(\cdot\,\rightarrow\cdot\,; C\right)1_{C}>
}\cr
&\times\tilde\rho^{\hbox{\srm geo}}\left(\un{\G}^{+}(\s_\L)\right)
\sum_{{K\sb \L;\diam(K)> r \hbox{ or }K=\em}\atop
{\s_{K}\neq \const}}
\r^{\hbox{\srm HT1}}\left(K, \s_{K},\eta_{K}\right)\cr
&\quad\times
\sum_{{K_1\sb \L;\diam(K_1)> r\hbox{ or }K_1=\em}
\atop{K_1\cap \del (\L)^c\neq \em} }
\r^{\hbox{\srm HT2}}\left(K_1, \s_{K_1},\eta_{K_1}\right)
}
\tag{4.27}
$$
which proves the desired representation (4.3)
with the obvious definition 
$$
\eqalign{
&\r_0(\G; \eta_{\un\G})\cr
&:=\r^{\hbox{\srm LT},\tilde m_{\del \L}}(\s_{\L})
e^{\frac{a m^*}{q}\sum_{{C\sb \L;\diam(C)\leq r}\atop
{\s_{C}\neq \const} }<\eta_C,
\RR\left(\cdot\,\rightarrow\cdot\,; C\right)\s_{C}>}\cr
&\times\tilde\rho^{\hbox{\srm geo}}\left(\un{\G}^{+}(\s_\L)\right)
\sum_{{K_0,K_1\sb \L; K_0\cup K_1\cup \un{\G}^{+}(\s_\L)=\un\G
;\diam(K_i)> r \hbox{ or }K_i=\em}\atop
{\s_{K_0}\neq \const, K_1\cap \del (\L)^c\neq \em}}
\r^{\hbox{\srm HT1}}\left(K_0, \s_{K_0},\eta_{K_0}\right)
\r^{\hbox{\srm HT2}}\left(K_1, \s_{K_1},\eta_{K_1}\right)\cr
}
\tag{4.28}
$$
The form (4.6) of the Peierls constant $\tilde \b_{\hbox{\srm Gau\ss}}$  
is now clear from $\tilde \b_{\hbox{\srm Gau\ss}}= \Const \min\{
\b(2r+1)^{-d},\a_0 \}-m^*\d$, assuming that 
both terms in the minimum are sufficiently 
large to control the entropy in (4.28)
and the slight modification in the exponential
bounds on $\r^{\hbox{\srm HT1}}$
arising from the subtraction of bounds. 
\endproof


\overfullrule=0pt
\bigskip\bigskip
\chap{V. The final contour model - Proof of phase transition}

We put together the results of Chapter III and IV
to obtain the contour representation of the full model.
It is of the same form as the Gaussian model
of Chapter IV, while a modifaction of the 
Peierls constant $\tilde\b$
accounts for the anharmonic contributions. More 
precisely we have

\proposition{5.1}{\it  Assume that the anharmonic $I$-weights (2.17)
satisfy the Positivity (2.19) 
and the uniform Peierls Condition
(2.20) with a constant $\e$. 
Suppose that $\e$
is sufficiently small, $q$ is sufficiently small, 
$a$ is of the order one,
$q(m^*)^2$ sufficiently large. 
Suppose that $\d\leq \Const m^*$ 
and $|U|\leq \Const m^*$
with constants that are sufficiently small.

Then the measures 
$T\left(\mu_{\L}^{+m^* 1_{\del \L},\eta_{\L}}\right)$
on $\{-1,1\}^{\L}$ have the contour representation
$$
\eqalign{
&T\left(\mu_{\L}^{+m^* 1_{\del \L},\eta_{\L}}\right)\left(\s_{\L}\right)\cr
&=\frac{1}{Z_{\hbox{\srm contour},\L}^{+,\eta_{\L}}}
e^{\sum_{C\sb V^+(\s_{\L})}S_{C}(\eta_C)
-\sum_{C\sb V^-(\s_{\L})}S_{C}(\eta_C)}
\sum_{{\G}\atop {\s_{\L}\left(\G\right)=\s_{\L}}}
\r(\G; \eta_{\un\G})\cr
}
\tag{5.1}
$$
with the contour-model partition function
$$
\eqalign{
&Z_{\hbox{\srm contour},\L}^{+,\eta_{\L}}
=\sum_{\G}
e^{\sum_{C\sb V^+(\G_{\L})}S_{C}(\eta_C)
-\sum_{C\sb V^-(\G_{\L})}S_{C}(\eta_C)}
\r(\G; \eta_{\un\G})\cr
}
\tag{5.2}
$$
For the partition function (1.4) we have 
$Z_{\L}^{+m^* 1_{\del \L},\eta_{\L}}
=C_{\L}^{+,\eta_{\L}}
Z_{\hbox{\srm contour},\L}^{+,\eta_{\L}}
$
with a trivial constant containing the contributions
of Gaussian fluctuations that satisfies, a.s. 
$$
\eqalign{
&\lim_{\L\uparrow Z^d} 
\frac{1}{|\L|}
\log C_{\L}^{+,\eta_{\L}}
\cr
&=-\frac{\E\eta^2_0}{2}
\left[\left(a-q\D_{\Z^d}   \right)^{-1}\right]_{0,0}
-\frac{1}{2}\left[\log\left( a-q\D_{\Z^d}\right)\right]_{0,0}
-b+\frac{1}{2}\log(2\pi)
}
\tag{5.3}
$$
The quantities appearing in (5.1) are as follows. 

\item{(i)} $\eta_C\mapsto S_{C}(\eta_C)$ 
are functions of the random fields indexed 
by the connected sets $C\sb \L$ that are symmetric, i.e.
$S_C(-\eta_C)=-S_C(\eta_C)$. 
In particular we have 
$S_x(\eta_x)=\frac{a m^*}{a+2dq }\eta_x$.
They obey the uniform bound
$$
\eqalign{
&\left|S_C(\eta_C)\right|\leq  m^* \d e^{-\a_{\srm final}|C|}
\cr
}
\tag{5.4}
$$
for all $C$
with $\a_{\srm final}=\const \min\left
\{\log\frac{1}{q},\log\frac{1}{\e} \left(\frac{\log \frac{1}{q}}{\log m^*}
\right)^d \right\}$.

\item{(ii)}
The activity $\r_{\hbox{\srm Ising}}(\G;\eta_{\un\G})$ 
is non-negative and depends only on the indicated arguments. 
It factorizes over the connected components (as in (4.4)).
For $\un\G$ not touching the boundary it does 
not depend on $\L$ and has
the infinite volume symmetries of a) invariance under 
joint flips of spins and random fields and
b) invariance under lattice shifts. 

There exist (large) positive constants $\tilde \b,\b$ s.t. 
we have the Peierls-type bounds:
$$
\eqalign{
&\r_{\hbox{\srm Ising}}(\G;\eta_{\un\G})
\leq
e^{-\b E^s(\G)-\tilde \b |\un \G|  }
\cr
}
\tag{5.5}
$$
uniformly in $\eta_G$.
Here $\b=\frac{q (m^*)^2}{2} \frac{a^2}{(a+2dq)^2-q^2}$
is the same as in (4.8)
and 
$$
\eqalign{
&\tilde \b=\Const \times \min\left\{
\log\frac{1}{q}, q{m^*}^2 \left(\frac{\log \frac{1}{q}}{\log m^*}
\right)^d,
\log\frac{1}{\e} \left(\frac{\log \frac{1}{q}}{\log m^*}
\right)^d\right\}-m^*\d
\cr
}
\tag{5.6}
$$
}

\proof Assuming the control 
of the anharmonicity, summarized in Proposition 3.1,
the proof is easy.
For any fixed $\s_{\L}$ we can cluster-expand the last sum
in (3.1). Dropping now the dependence
on the boundary condition $\tilde m_{\del \L}=+m^* 1_{\del \L}$
in the notation we have
$$
\eqalign{
&\log \sum_{G:\em\sb G \subset \L } 
\bar\r\left(G; \s_{G}
,\eta_{G} \right)
=\sum_{C:\em\sb C \subset \L } 
\bar\e\left(C; \s_{C}
,\eta_{C} \right)\cr
&=\sum_{C\sb V^+(\s_{\L})}\bar\e\left(C;1_{C}
,\eta_{C} \right)
+\sum_{C\sb V^-(\s_{\L})}\bar\e
\left(C;-1_{C},\eta_{C} \right)
+\sum_{C\sb \L;\s_C\neq \const}
\bar\e\left(C; \s_{C},\eta_{C} \right)\cr
\cr
}
\tag{5.7}
$$
where the sum is over connected sets $C$
and we have the bounds $\left|\e\left(C; \s_{C},\eta_{C} \right)\right|
\leq e^{-\const \a|C|}$
with $\a$ given in Proposition 3.1.
Together with the representation (4.3) for the 
purely Gaussian model this gives
$$
\eqalign{
&e^{-\inf_{m_{\L}\in \R^{\L}} H^{\tilde m_{\del \L},\eta_{\L}, 
\s_\L}_{\L}\left(m_{\L}\right)}
\sum_{G:\em\sb G \subset \L } \bar\r\left(G; \s_{G}
,\eta_{G} \right)
=K_{\L}\left(\eta_{\L}\right)\cr
&\times e^{\sum_{C\sb V^+(\s_{\L})}\left(S^{\hbox{\srm Gau\ss}}_{C}(\eta_C)
+\bar\e\left(C;1_{C},\eta_{C} \right)\right)
-\sum_{C\sb V^-(\s_{\L})}\left(S^{\hbox{\srm Gau\ss}}_{C}(\eta_C)
+\bar\e\left(C;1_{C}
,\eta_{C} \right)\right)
}\cr
&\times 
e^{\sum_{C\sb \L;\s_C\neq \const}
\bar\e\left(C; \s_{C}
,\eta_{C} \right)}
\sum_{{\G}\atop {\s_{\L}\left(\G\right)=\s_{\L}}}
\r_0(\G; \eta_{\un\G})\cr
}
\tag{5.8}
$$
Note that the $C$'s in the exponential
in the last line are in particular connected to $\G$.
Using subtraction-of-bounds as before we can expand
those terms and, as we did before in Chapter III and IV, 
rewrite the last line in terms
of a new (and final) contour summation as 
$$
\eqalign{
&e^{\sum_{C\sb \L;\s_C\neq \const}
\bar\e\left(C; \s_{C}
,\eta_{C} \right)}
\sum_{{\G}\atop {\s_{\L}\left(\G\right)=\s_{\L}}}
\r_0(\G; \eta_{\un\G})
=\sum_{{\G}\atop {\s_{\L}\left(\G\right)=\s_{\L}}}
\r(\G; \eta_{\un\G})\cr
}
\tag{5.9}
$$
The values of the Peierls constants for the final
activities on the r.h.s. 
follow from the statements of the Propositions 3.1 and 4.1
with a slight loss due to the control of 
entropy. 

Finally, to see the statement for the free energy, 
we start from (3.1) and recall the construction
of the activities in the purely Gaussian case, starting 
from (4.16). Using the explicit expression (4.11)
for the energy minimum in the Gaussian model in terms
of the resolvent we obtain, with some trivial 
control on boundary terms, using the SLLN applied on the random
fields the desired formula 
$$
\eqalign{
&\lim_{\L\uparrow Z^d} 
\frac{1}{|\L|}
\log C_{\L}^{+,\eta_{\L}}
\cr
&=-\lim_{\L\uparrow Z^d} 
\frac{1}{|\L|}\E\inf_{m_{\L}\in \R^{\L}} H^{\tilde m_{\del \L},\eta_{\L}, 
1_\L}_{\L}\left(m_{\L}\right)
-\lim_{\L\uparrow Z^d}\frac{1}{2|\L|}\log\det\left( a-q\D_{\L}\right)
-b+\frac{1}{2}\log(2\pi)\cr
&=-\frac{\E\eta^2_0}{2}
\left[\left(a-q\D_{\Z^d}   \right)^{-1}\right]_{0,0}
-\frac{1}{2}\left[\log\left( a-q\D_{\Z^d}\right)\right]_{0,0}
-b+\frac{1}{2}\log(2\pi)
}
\tag{5.10}
$$
\endproof

The following result provides control of the original
measure in terms of the coarse-grained one
up to two corrections:

\proposition{5.2}{\it 
Assume the conditions of Proposition 5.1
and suppose that $\tilde m_{\del \L}\in \left(U^+\right)^{\del \L}$.
Then we have 
$$
\eqalign{
&\mu_{\L}^{\tilde m_{\del \L},\eta_{\L}}\left[m_{x_0}\leq \frac{m^*}{2}\right]
\leq \left(T\left(\mu_{\L}^{\tilde m_{\del \L},\eta_{\L}}\right)\right)
\left[\s_{x_0}=-1
\right]+e^{-\const\a}+e^{-\const (m^*)^2}\cr
\cr
}
\tag{5.11}
$$
where $\a=\const \times \min\left\{
\log\frac{1}{q},\log\frac{1}{\e} \left(\frac{\log \frac{1}{q}}{\log m^*}
\right)^d\right\}$ is given in Proposition 3.1.\
}

\remark The first term on the r.h.s. 
accounts for the anharmonicity, the 
next one for the Gaussian fluctuations.

\proof
We carry out the transformation that
led to Lemma 2.3 while 
carrying through the indicator function $1_{m_{x_0}\leq \frac{m^*}{2}}$
to get 
$$
\eqalign{
&\int_{\R^{\L}}dm_{\L}1_{m_{x_0}\leq \frac{m^*}{2}}
e^{-E^{\tilde m_{\del \L},\eta_{\L}}_{\L}
\left(m_{\L}\right)}=
e^{-b|\L|}
\sum_{\s_\L}e^{-\inf_{m'_{\L}} H^{\tilde m_{\del \L},\eta_{\L}, 
\s_\L}_{\L}\left(m'_{\L}\right)}\cr
&\times \Biggl[\sum_{{G:G \subset \L}\atop{G\ni x_0} }\left(2\pi \right)^{\frac{|\L\ba\ov{G}|}{2}}
\left(\det \left(a-q \D_{\L\ba\ov G} \right)\right)^{-\frac{1}{2}}
\int dm_{\del G}
e^{-\D H^{\tilde m_{\del \L},
\eta_{\L},\s_{\L}}_{\del G,\L}\left(m_{\del G} \right)}
1_{m_{\del G}\in U^{\del G}}I^{\tilde m_{\del \L},m_{\del G},
\eta_{G},\s_{G}}_{G;x_0}\cr
&+\sum_{{G:G \subset \L}\atop{\del G\ni x_0} }
\left(2\pi \right)^{\frac{|\L\ba\ov{G}|}{2}}
\left(\det \left(a-q \D_{\L\ba\ov G} \right)\right)^{-\frac{1}{2}}
\int dm_{\del G}
e^{-\D H^{\tilde m_{\del \L},
\eta_{\L},\s_{\L}}_{\del G,\L}\left(m_{\del G} \right)}
1_{m_{\del G}\in U^{\del G}}1_{m_{x_0}\leq \frac{m^*}{2}}I^{\tilde m_{\del \L},m_{\del G},
\eta_{G},\s_{G}}_{G}\cr
&+
\sum_{{G:G \subset \L}\atop{\ov G\not\ni x_0} }
\int dm_{\del G}
e^{-\D H^{\tilde m_{\del \L},
\eta_{\L},\s_{\L}}_{\del G,\L}\left(m_{\del G} \right)}
1_{m_{\del G}\in U^{\del G}}
\int dm_{\L\ba\ov G}e^{
-\D H^{\tilde m_{\del \L},m_{\del G},
\eta_{\L\ba\ov{G}},\s_{\L\ba\ov{G}}}
_{\L\ba\ov{G}}\left(m_{\L\ba\ov{G}} \right)}1_{m_{x_0}\leq \frac{m^*}{2}}\cr
&\quad\times I^{\tilde m_{\del \L},m_{\del G},
\eta_{G},\s_{G}}_{G}
\Biggr]
}
\tag{5.12}
$$
with  $I_{G;x_0}
=I^{(1)}_{G;x_0}-I^{(2)}_{G;x_0}$ (superscripts are dropped now)
where we have defined 
$$
\eqalign{
&I^{(1)}_{G;x_0}
:=\int dm_{G}e^{-\D H^{\tilde m_{\del \L},m_{\del G},
\eta_{G},\s_{G}}_{G}\left(m_{G} \right)}
\prod_{x\in G}\left(1_{m_x\not\in U}+w(m_x) \right)1_{m_{x_0}\leq \frac{m^*}{2}}
\cr
&I^{(2)}_{G;x_0}
:=\int dm_{G}e^{-\D H^{\tilde m_{\del \L},m_{\del G},
\eta_{G},\s_{G}}_{G}\left(m_{G} \right)}
\prod_{x\in G}1_{m_x\not\in U}\,\,\,
1_{m_{x_0}\leq \frac{m^*}{2}}\cr
}
\tag{5.13}
$$
We use the same notations {\it without} the subscript $x_0$
on the l.h.s. to denote the integrals without the $1_{m_{x_0}\leq \frac{m^*}{2}}$
on the r.h.s. so that we have $I_{G}=I^{1}_{G}-I^{2}_{G}$.
Note that it is not clear anymore that 
$I_{G;x_0}$ is positive 
for any sign $\s_{x_0}$ and dominated by $I_{G}$.
To bypass this little inconvenience we argue as follows. 
Let us slightly enlarge $b$ in Chapter II by putting a 
factor $2$ in front of the fraction of integrals 
in the definition (2.42). 
This leaves $b$ very small and all
subsequent arguments based on a fixed choice of $b$
remain valid. 
Going back through Lemma 2.4. we see that this 
definition implies 
that even $I^{(2)}_{G}\leq 2^{-|G|}I^{(1)}_{G}$
(which can be seen as a strengthening of the positivity of $I_{G}$).
But from this we have in particular that 
$$
\eqalign{
&I_{G;x_0}
=I_{G;x_0}^{(1)}-I_{G;x_0}^{(2)}
\leq I_{G;x_0}^{(1)}\leq I_{G}^{(1)}\leq 2 I_{G}
}
\tag{5.14}
$$
We use this estimate on the last $G$-sum in (5.12)
and bound the second  $G$-sum in (5.12)
by the corresponding expression without the indicator. 
Carrying out the $m_{\del G}$-integral
as described in Chapter III we get from this the bound
$$
\eqalign{
&\sum_{{G:G \subset \L}\atop{G\ni x_0} }\left(2\pi \right)^{\frac{|\L\ba\ov{G}|}{2}}
\left(\det \left(a-q \D_{\L\ba\ov G} \right)\right)^{-\frac{1}{2}}
\int dm_{\del G}
e^{-\D H^{\tilde m_{\del \L},
\eta_{\L},\s_{\L}}_{\del G,\L}\left(m_{\del G} \right)}
1_{m_{\del G}\in U^{\del G}}I^{\tilde m_{\del \L},m_{\del G},
\eta_{G},\s_{G}}_{G;x_0}\cr
&+\sum_{{G:G \subset \L}\atop{\del G\ni x_0} }
\left(2\pi \right)^{\frac{|\L\ba\ov{G}|}{2}}
\left(\det \left(a-q \D_{\L\ba\ov G} \right)\right)^{-\frac{1}{2}}
\int dm_{\del G}
e^{-\D H^{\tilde m_{\del \L},
\eta_{\L},\s_{\L}}_{\del G,\L}\left(m_{\del G} \right)}
1_{m_{\del G}\in U^{\del G}}1_{m_{x_0}\leq \frac{m^*}{2}}I^{\tilde m_{\del \L},m_{\del G},
\eta_{G},\s_{G}}_{G}\cr
&\leq
2\cdot\left(2\pi \right)^{\frac{|\L|}{2}}
\left(\det \left(a-q \D_{\L} \right)\right)^{-\frac{1}{2}}
\sum_{G:x_0\in \ov {G} \subset \L } 
\bar\r^{\tilde m_{\del_{\del \L} G}}\left(G; \s_{G}
,\eta_{G} \right)
}
\tag{5.15}
$$ 
Using the positivity of the activities
in the last line we can use the usual Peierls
argument on the fixed-$\s$ contour model appearing in (3.1)
that controls the anharmonicity.
So we estimate
$$
\eqalign{
\sum_{G:x_0\in \ov{G} \subset \L } 
\bar\r^{\tilde m_{\del_{\del \L} G}}\left(G; \s_{G}
,\eta_{G}\right) 
&\leq \sum_{G_0:x_0\in \ov{G_0} \subset \L } 
\bar\r^{\tilde m_{\del_{\del \L} G}}\left(G_0; \s_{G_0}
,\eta_{G_0} \right)
\sum_{G: G \subset \L } 
\bar\r^{\tilde m_{\del_{\del \L} G}}\left(G; \s_{G}
,\eta_{G}\right)\cr
&\leq e^{-\const \a}\sum_{G: G \subset \L } 
\bar\r^{\tilde m_{\del_{\del \L} G}}\left(G; \s_{G}
,\eta_{G}\right)
}
\tag{5.16}
$$ 
where the first sum is over connected sets $G_0$
and we have used Proposition (3.1) for its estimation. 

To treat the first $G$-sum in (5.12) we note that the expectation
outside the anharmonic contours is given by the one-dimensional
Gaussian probability, 
$$
\eqalign{
&\int dm_{\del G}
e^{-\D H^{\tilde m_{\del \L},
\eta_{\L},\s_{\L}}_{\del G,\L}\left(m_{\del G} \right)}
1_{m_{x_0}\leq \frac{m^*}{2}}
=\left(2\pi \right)^{\frac{|\L\ba\ov{G}|}{2}}
\left(\det \left(a-q \D_{\L\ba\ov G} \right)\right)^{-\frac{1}{2}}
\cr
&\times \NN\left[m_{\L\ba\del G}^{\tilde m_{\del \L},m_{\del G},
\eta_{\L\ba\ov G},
\s_{\L\ba\ov G}};
\left(a-q\D^{D}_{\L-\ov{G}}\right)^{-1}_{x_0,x_0}
\right](1_{m_{x_0}\leq \frac{m^*}{2}})
}
\tag{5.17}
$$
with the notation 
$\NN[a;\s^2](\phi)=\int_{-\infty}^\infty\frac{
e^{-\frac{(x-a)^2}{2\s^2}}\phi(x)
}{\sqrt{2\pi\s^2}} 
$.
We use the uniform control on the expectation value
given by Lemma 2.5 and the fact that 
the variance occuring in (5.17) 
is of the order one, in any volume.
If $\s_{x_0}=+1$ we have from this, uniformly in all
involved quantities that
$$
\eqalign{
&\NN\left[m_{\L\ba\del G}^{\tilde m_{\del \L},m_{\del G},
\eta_{\L\ba\ov G},
\s_{\L\ba\ov G}};
\left(a-q\D^{D}_{\L-\ov{G}}\right)^{-1}_{x_0,x_0}
\right](1_{m_{x_0}\leq \frac{m^*}{2}})\leq e^{-\const (m^*)^2}
}
\tag{5.18}
$$
so it can be pulled out of the $m_{\del G}$-integral.
For $\s_{x_0}=-1$ we use the trivial bound $1$ to write
$$
\eqalign{
&\sum_{{G:G \subset \L}\atop{\ov G\not\ni x_0} }\int dm_{\del G}
e^{-\D H^{\tilde m_{\del \L},
\eta_{\L},\s_{\L}}_{\del G,\L}\left(m_{\del G} \right)}
1_{m_{\del G}\in U^{\del G}}
\int dm_{\L\ba\ov G}e^{
-\D H^{\tilde m_{\del \L},m_{\del G},
\eta_{\L\ba\ov{G}},\s_{\L\ba\ov{G}}}
_{\L\ba\ov{G}}\left(m_{\L\ba\ov{G}} \right)}1_{m_{x_0}\leq \frac{m^*}{2}}\cr
&\leq 
\left(
e^{-\const (m^*)^2}1_{\s_{x_0}=1}+1_{\s_{x_0}=-1}
\right)\cr
&
\sum_{{G:G \subset \L}\atop{\ov G\not\ni x_0} }
\left(2\pi \right)^{\frac{|\L\ba\ov{G}|}{2}}
\left(\det \left(a-q \D_{\L\ba\ov G} \right)\right)^{-\frac{1}{2}}
\int dm_{\del G}
e^{-\D H^{\tilde m_{\del \L},
\eta_{\L},\s_{\L}}_{\del G,\L}\left(m_{\del G} \right)}
1_{m_{\del G}\in U^{\del G}}I^{\tilde m_{\del \L},m_{\del G},
\eta_{G},\s_{G}}_{G}
\cr
&\leq 
\left(
e^{-\const (m^*)^2}+1_{\s_{x_0}=-1}
\right)\cr
&
\sum_{G:G \subset \L }
\left(2\pi \right)^{\frac{|\L\ba\ov{G}|}{2}}
\left(\det \left(a-q \D_{\L\ba\ov G} \right)\right)^{-\frac{1}{2}}
\int dm_{\del G}
e^{-\D H^{\tilde m_{\del \L},
\eta_{\L},\s_{\L}}_{\del G,\L}\left(m_{\del G} \right)}
1_{m_{\del G}\in U^{\del G}}I^{\tilde m_{\del \L},m_{\del G},
\eta_{G},\s_{G}}_{G}
\cr
}
\tag{5.19}
$$
Now it is simple to put together (5.12), (5.16)-(5.19) 
and rerunning the next steps of the transformation 
yields the claim.\endproof


Applying the information of [BK1] 
we obtain the main result of the paper. 

\proofof{Theorem 1}
We apply statement Theorem 2.1 [BK1]
on the measure $T\left(\mu_{\L}^{+m^* 1_{\del \L},\eta_{\L}}\right)$.
Indeed, this is justified from Proposition 5.1
which implies that this measure
is contained in the class of contour measures
described in [BK1] Chapter 5 `Flow of the RGT', Paragraph 5.1.
We note that of the three constants $\tilde \b$, $\b$, 
$\a_{\srm final}$ 
(controlling
the exponential decay of the activities
in terms of the volume resp. in terms of the 
naive contour energy, and the decay of the non-local
fields) the constant $\tilde \b$
is the smallest.

So statement (2.3) from [BK1] gives in our case that
for $d\geq 3$, $\tilde \b$  large enough and
$\s^2$ small enough we have that 
$$
\eqalign{
&\P\left[
T\left(\mu_{\L}^{+m^* 1_{\del \L},\eta_{\L}}\right)
\left[\s_{x_0}=-1\right]
\geq e^{-\const \tilde \b}\right]
\leq  e^{-\frac{\const}{\s^2}} 
\cr
}
\tag{5.20}
$$
We apply our Proposition (5.2) 
and note that the two correction
terms given therein are also controlled by $e^{-\const \tilde\b}$
(with possible modification of $\const$.)
From this in particular the estimates of Theorem 1 follow.

\endproof

\remark We have not given an estimate on the value of $\g$
as a function of $q$ and $m^*$.
This would of course follow from a more careful estimate 
of the best value of the `anharmonicity-constant' 
$\e$ (which is entering $\tilde \b$) as a function of $q$ and $m^*$
(see Chapter II) and is left to the reader.

\bigskip
\bigskip

Finally, Theorem 2 for the $\phi^4$-theory
follows immediately from 

\proposition{5.3}{\it  Assume that the anharmonic $I$-weights (2.17)
satisfy the Positivity (2.19) 
and the uniform Peierls Condition
(2.20) with a constant $\e$ (that is sufficiently small). 
Let $\mu^\eta_{\infty}$ be any continuous spin Gibbs-measure 
obtained as a weak limit of 
$\mu_{\L}^{\tilde m_{\del \L},\eta_{\L}}$
along  a sequence of cubes $\L$ 
for some 
(not necessarily positive) continuous-spin boundary condition 
$\tilde m \in U^{\Z^d}$.

Then the measure
$T\left(\mu^\eta_{\infty}\right)$
on $\{-1,1\}^{\Z^d}$ is a Gibbs measure
for the absolutely summable Ising-Hamiltonian
$$
\eqalign{
&H^{\eta}_{\srm Ising}\left(\s\right)\cr
&=-\frac{a^2 (m^*)^2}{2}\sum_{x,y}\left( 
a-q\D_{\Z^d}\right)^{-1}_{x,y}\s_{x}\s_{y}
- a m^* \sum_{x}\left(a-q\D_{\Z^d}\right)^{-1}_{x,y}\eta_x \s_x
+\sum_{C:|C|\geq 2}\Phi_{C}\left(\s_C;\eta_C\right)
}
\tag{5.21}
$$
where the interaction potentials
$\Phi_C(\s_C,\eta_{C})=\Phi_C(-\s_C,-\eta_{C})$
obey the uniform bound
$\left|\Phi_C(\s_{C},\eta_C)\right|\leq e^{-\const \a|C|}$
for all $C$ with
$\a=\const \times \min\left\{
\log\frac{1}{q},\log\frac{1}{\e} \left(\frac{\log \frac{1}{q}}{\log m^*}
\right)^d\right\}$ as in (3.3).
}

\noindent {\bf Remark: }Note that it follows 
in particular that the interaction
will be the same e.g. also in continuous spin 
Dobrushin-states [Do2] (that are believed
to exist) one  could construct using the boundary
condition $+m^*$ in the upper half-space and 
$-m^*$ in the lower half-space.

\proof Denote by $H_{\srm Ising,V}^{\bar \s_{\Z^d}}
\left(\s_V\right)$ the usual restriction 
of (5.21) to the finite volume $V$, 
obtained by keeping the sums over 
sets $\{x,y\}$ and $C$ that intersect $V$ 
and putting the
spin equal to $\bar \s_{\Z^d}$ for $x\not\in V$.
Following [BKL] it suffices to 
show that, for each $\bar \s_{\Z^d}$ we have that
$$
\eqalign{
&\lim_{\L_2\uparrow \Z^d}
\lim_{\L_1\uparrow \Z^d}\frac{Z^{\tilde m_{\del\L_1},\eta_{\L_1}}_{\L_1}
\left(  
\s_{V}, \bar\s_{\L_2\ba V}
\right)}{
\sum_{\tilde \s_V}
Z^{\tilde m_{\del\L_1},\eta_{\L_1}}_{\L_1}\left(  
\tilde\s_{V}, \bar\s_{\L_2\ba V}
\right)}
=\frac{e^{-H_{\srm Ising,V}^{\bar \s_{\Z^d},\eta}
\left(\s_V\right) }}
{\sum_{\tilde\s_V} e^{-H_{\srm Ising,V}^{\bar \s_{\Z^d},\eta}
\left(\tilde \s_V\right) }}
}
\tag{5.22}
$$
along (say) sequences of cubes where 
$$
\eqalign{
&Z^{\tilde m_{\del\L_1},\eta_{\L_1}}_{\L_1}\left(  
\s_{\L_2}\right)
:=\int_{\R^{\L_1}}
e^{-E^{\tilde m_{\del \L_1},\eta_{\L_1}}_{\L_1}
\left(m_{\L_1}\right)}\prod_{x\in \L_2}
T_x\left(\s_x\bigl| m_x\right)
\cr
}
\tag{5.23}
$$
This is clear, since (according to our 
assumption of weak convergence) 
there is a subsequence of cubes $\L_1$
s.t. the inner limit exists and equals 
$\left(T\left(\mu^\eta_{\infty}\right)
\right)\bigl(\s_{V}\bigl|\bar \s_{\L_2\ba V} \bigr)$.
Summing Proposition 3.1 over the spins in $\L_1\ba \L_2$
we have then
$$
\eqalign{
&Z^{\tilde m_{\del\L_1},\eta_{\L_1}}_{\L_1}\left(  
\s_{\L_2}\right)
=e^{-b|\L_1|}\left(2\pi \right)^{\frac{|\L_1|}{2}}
\left(\det \left(a-q \D_{\L_1} \right)\right)^{-\frac{1}{2}}
\sum_{\hat\s_{\L_1\ba \L_2}}
e^{-\inf_{m'_{\L_1}} H^{\tilde m_{\del \L_1},\eta_{\L_1}, 
\hat\s_{\L_1\ba \L_2}; \s_{\L_2}}_{\L_1}\left(m'_{\L_1}\right)}\cr
&\times\sum_{G:\em\sb G \subset \L_1 } 
\bar\r^{\tilde m_{\del_{\del \L_1} G}}\left(G;\s_{G\cap \L_2},
\hat\s_{G\cap \L_1\ba \L_2}
,\eta_{G} \right)\cr
}
\tag{5.24}
$$
From here the proof is easy, given the
explicit formula (4.11) for the minimum and
the absolute summability of the polymer weights,
uniformly in the spins and random fields.
 
For the convenience of the reader
we give a complete proof for the simplest case 
of vanishing anharmonicity $w(m_x)\equiv 0$,
and vanishing magnetic fields 
$\eta_x=0$; it illustrates the way boundary terms
are entering. 
Using (4.11) we have indeed
$$
\eqalign{
&Z^{\tilde m_{\del\L_1},\eta_{\L_1}}_{\L_1}\left(  
\s_{\L_2}\right)\cr
&:=\Const \times 
\sum_{\s_{\L_1\ba \L_2}}
e^{\frac{ a^2 (m^*)^2}{2 q}<\left(\s_{\L_2},\s_{\L_1\ba \L_2}\right),
R_{\L_1}\left(\s_{\L_2},\s_{\L_1\ba \L_2}\right)     >_{\L_1}
+a m^*<\tilde\eta_{\del({\L_1}^c)}(\tilde m),R_{\L_1}
\left(\s_{\L_2},\s_{\L_1\ba \L_2}\right)>_{\L_1}
}
}
\tag{5.25}
$$
Now, using the exponential decay of the resolvent, 
$$
\eqalign{
&Z^{\tilde m_{\del\L_1},\eta_{\L_1}}_{\L_1}\left(  
\s_{V}, \bar\s_{\L_2\ba V}
\right)\cr
&=\Const 
\times 
e^{\frac{ a^2 (m^*)^2}{2 q}<\left(\s_{V},\bar\s_{\L_2\ba V}\right),
R_{\L_1}\left(\s_{V},\bar\s_{\L_2\ba V}\right)
>_{\L_1}
+a m^*<\tilde\eta_{\del({\L_1}^c)}(\tilde m),R_{\L_1}
\left(\s_{V},\bar\s_{\L_2\ba V}\right)>_{\L_1}
}\cr
&\times\sum_{\s_{\L_1\ba \L_2}}
e^{\frac{ a^2 (m^*)^2}{q}<
\left(\s_{V},\bar\s_{\L_2\ba V}\right),
R_{\L_1}\s_{\L_1\ba \L_2}     >_{\L_1}
+\frac{ a^2 (m^*)^2}{2 q}<\s_{\L_1\ba \L_2},
R_{\L_1}\s_{\L_1\ba \L_2}     >_{\L_1}
+a m^*<\tilde\eta_{\del({\L_1}^c)}(\tilde m),R_{\L_1}
\s_{\L_1\ba \L_2}>_{\L_1}
}\cr
&=\Const 
\times 
e^{\frac{ a^2 (m^*)^2}{2 q}<\left(\s_{V},\bar\s_{\L_2\ba V}\right),
R_{\L_1}\left(\s_{V},\bar\s_{\L_2\ba V}\right)
>_{\L_1}
\pm \Const |\L_2|e^{-\a'\hbox{\srm dist}(\L_2,\L_1^c)}
}\times
e^{\pm\Const |V|e^{-\a'\hbox{\srm dist}(V,\L_2^c)}}
\cr
&\times \sum_{\s_{\L_1\ba \L_2}}e^{\frac{ a^2 (m^*)^2}{q}<\bar\s_{\L_2\ba V},
R_{\L_1}\s_{\L_1\ba \L_2}     >_{\L_1}
}
e^{+\frac{ a^2 (m^*)^2}{2 q}<\s_{\L_1\ba \L_2},
R_{\L_1}\s_{\L_1\ba \L_2}     >_{\L_1}
+a m^*<\tilde\eta_{\del({\L_1}^c)}(\tilde m),R_{\L_1}
\s_{\L_1\ba \L_2}>_{\L_1}
}
}
\tag{5.26}
$$
The terms in the last sum do not depend 
on $\s_{V}$ so that we get
$$
\eqalign{
&\frac{Z^{\tilde m_{\del\L_1},\eta_{\L_1}}_{\L_1}\left(  
\s_{V}, \bar\s_{\L_2\ba V}
\right)}{
\sum_{\tilde \s_V}
Z^{\tilde m_{\del\L_1},\eta_{\L_1}}_{\L_1}\left(  
\tilde\s_{V}, \bar\s_{\L_2\ba V}
\right)}\cr
&=e^{\pm \Const |\L_2|e^{-\a'\hbox{\srm dist}(\L_2,\L_1^c)}
\pm\Const |V|e^{-\a'\hbox{\srm dist}(V,\L_2^c)}}
\frac{e^{\frac{ a^2 (m^*)^2}{2 q}<\left(\s_{V},\bar\s_{\L_2\ba V}\right),
R_{\L_1}\left(\s_{V},\bar\s_{\L_2\ba V}\right)
>_{\L_1}}}
{\sum_{\tilde \s_{V}}
e^{\frac{ a^2 (m^*)^2}{2 q}<\left(\tilde \s_{V},\bar\s_{\L_2\ba V}\right),
R_{\L_1}\left(\tilde \s_{V},\bar\s_{\L_2\ba V}\right)
>_{\L_1}}}\cr
}
\tag{5.27}
$$
with uniform constants.
Taking first $\L_1\uparrow \Z^d$ (using that $R_{\L_1}\bigl|_{\L_2}\rightarrow
R_{\Z^d}\bigl|_{\L_2} $)
and then $\L_2\uparrow \Z^d$ we get in fact
the desired result in our special case. 

The (random) non-Gaussian case follows easily from 
the cluster expansion of the $G$-sum in (3.1).
Indeed, since we have uniform exponential
decay of the activities $\bar\r^{\tilde m_{\del_{\del \L} G}}$,
the cluster expansion gives
us quantities
$\Phi_{C}^{\tilde m_{\del_{\del \L} C}}\left(\s_C;\eta_C\right)$ 
that obey a uniform bound of the form as desired 
s.t. we have 
$$
\eqalign{
&\sum_{G:\em\sb G \subset \L } 
\bar\r^{\tilde m_{\del_{\del \L} G}}\left(G; \s_{G}
,\eta_{G} \right)
= e^{\sum_{C\hbox{ \srm conn.}:\em\sb C \subset \L } 
\Phi_{C}^{\tilde m_{\del_{\del \L} C}}\left(\s_C;\eta_C\right)     } 
\cr
}
\tag{5.28}
$$
The Gibbs potential in (5.21) is then given by the value 
of $\Phi_{C}$ for polymers $C$ that are not touching
the boundary. 
With estimates on boundary terms
as in (5.26) the claim (5.22) follows.\endproof



\overfullrule=0pt
\bigskip\bigskip
\vfill\eject

\chap{Appendix}
For easy reference we collect some formulae
about quadratic forms and the random walk expansion
of determinants and correlation functions we use.
We start with

\lemma{A.1}{\it Let $Q_{\L}$ be symmetric and positive definite.
Let $V\sb \L$ and write, with obvious notations, 
$$
\eqalign{
&Q_{\L}=\left(\matrix{
  Q_{V}       &   Q_{V,\L\ba V}  \cr
  Q_{\L\ba V,V}     &   Q_{\L\ba V,\L\ba V}
}\right)\cr
}
\tag{A.1}
$$
Then we have the following formulae.

\item{(i)}
$$
\eqalign{
&Q_{\L}^{-1}=\cr
&\left(\matrix{
  \left(Q_{V}-Q_{V,\L\ba V}Q_{\L\ba V,\L\ba V}^{-1}Q_{\L\ba V,V}
\right)^{-1}     
& -Q_{V}^{-1}\left( Q_{\L\ba V,\L\ba V}- Q_{\L\ba V,V}Q_{V}^{-1} Q_{V,\L\ba V}\right)^{-1} \cr
- Q_{\L\ba V,\L\ba V}^{-1}\left(Q_{V}-Q_{V,\L\ba V}Q_{\L\ba V,\L\ba V}^{-1}Q_{\L\ba V,V}
\right)^{-1}       &   
\left( Q_{\L\ba V,\L\ba V}-Q_{\L\ba V,V}Q_{V}^{-1} Q_{V,\L\ba V}\right)^{-1}
}\right)\cr
}
\tag{A.2}
$$
\item{(ii)}
$$
\eqalign{
&\det Q_{\L}=\det \left(\Pi_{V}Q_{\L}^{-1}\Pi_{V}\right)^{-1}
\times \det Q_{\L\ba V}\cr
}
\tag{A.3}
$$
\item{(iii)}
For any $z_{\L}$ we can write
$$
\eqalign{
&\frac{1}{2}<m_\L,Q_{\L}m_\L>_{\L}
-<m_\L,z_\L>_{\L}\cr
&=\frac{1}{2}<\left(m_{V}-m^{z_{\L}}_{\L}\Bigl|_{V}\right)
,\left(\Pi_{V}Q_{\L}^{-1}\Pi_{V}\right)^{-1}
\left(m_{V}-m^{z_{\L}}_{\L}\Bigl|_{V}\right)>_{\L}\cr
&+\frac{1}{2}<\left(m_{\L-V}-
m^{z_{\L-V},m_{V}}_{\L\ba V}\right)
,Q_{\L\ba V}
\left(m_{V}-m^{z_{\L\ba V},m_{V}}_{\L\ba V}\right)>_{\L}
-\frac{1}{2}<z_\L,Q_{\L}^{-1}z_\L>_{\L}
}
\tag{A.4}
$$
where the `global minimizer' 
$m^{z_{\L}}_{\L}
=Q_{\L}^{-1}z_\L$
is the minimizer of the total energy, i.e. 
$$
\eqalign{
&m_{\L}\mapsto \frac{1}{2}<m_\L,Q_{\L}m_\L>_{\L}
-\frac{1}{2}<m_\L,z_\L>_{\L}\cr
}
\tag{A.5}
$$
We write $Q_{V}=\Pi_{V}Q_{\L}\Pi_{V}$,
$Q_{\L\ba V,V}=\Pi_{\L\ba V}Q_{\L}\Pi_{V}$. 
The `conditional
minimizer'
$$
\eqalign{
&m^{z_{\L\ba V},m_{V}}_{\L\ba V}
=Q_{\L\ba V}^{-1}\left(z_{\L\ba V} + Q_{\L\ba V,V}m_{V} \right)\cr
}
\tag{A.6}
$$
is the minimizer of the function
$$
\eqalign{
&m_{\L\ba V}\mapsto \frac{1}{2}<\left(m_{\L\ba V},m_{V}\right),
Q_{\L}\left(m_{\L\ba V},m_{V}\right)>_{\L}
-\frac{1}{2}<\left(m_{\L\ba V},m_{V}\right),z_\L>_{\L}\cr
}
\tag{A.7}
$$
for fixed $m_{V}$.

}

\noindent\remark The quadratic forms
on the diagonal of the r.h.s. of (A.2)
are automatically positive definite. 

The proofs are easy and well known computations
and will not be given here.
Next we collect some formulae and introduce notation
concerning the random walk representation.

\lemma{A.2}{\it Denote by 
$\RR$ the (non-normalized) measure on the set of all finite 
paths on $\Z^d$ (with all possible lengths),
defined by
$$
\eqalign{
&\RR\left(\{\g\}\right)=
\left(\frac{1}{c+2d}\right)^{|\g|+1}
}
\tag{A.8}
$$
for a nearest neighbor path $\g$ of finite length $|\g|$.
Then we have
$$
\eqalign{
&R_{\L;x,y}=\RR\left(
\g\hbox{ from }x\hbox{ to }y;
\hbox{ Range}(\g)\sb \L
\right)\cr
}
\tag{A.9}
$$
where  $\hbox{ Range}(\g)=\{\g_t;t=0,\dots,k\}$
is set of sites  
visited by a path $\g=(\g_t)_{t=0,\dots,k}$ of length $|\g|=k$.
}

\proof 
Write $\D^{D}_\L= 2d -T_{\L}$
where $T_{\L;x,y}=1$
iff $x,y\in \L$ are nearest neighbors and
$T_{\L;x,y}=0$ otherwise. Then 
$$
\eqalign{
&R_{\L;x,y}=
\left(c+2d -T_{\L}  \right)^{-1}_{x,y}
=\sum_{t=0}^\infty
\left(\frac{1}{c+2d}\right)^{t+1}
 \left(T_\L^{t}\right)_{x,y}\cr
}
\tag{A.10}
$$
which proves (A.10).
\endproof

We will also use the obvious matrix notation
$$
\eqalign{
&\left(\RR\left(\cdot\,\rightarrow\cdot\,; C\right)\right)_{x,y}
= \RR\left(
\g\hbox{ from }x \hbox{ to }y;
\hbox{ Range}(\g)=C\right)\cr
}
\tag{A.11}
$$
so that one has the matrix equality
$R_{V}=\sum_{C\sb V}
\RR\left(\cdot\,\rightarrow\cdot\,; C\right)$
for any volume $V$.
We need to use a bound on its matrix
elements at several places. Let us  note the simple 
estimate
$$
\eqalign{
&\sum_{y\in \Z^d}\RR\left(x\,\rightarrow y\,; C\right)
\leq \RR\left(
\g\hbox{ starting at }x,\hbox{ length}(\g)\geq |C|-1
\right)
=\frac{1}{c}\left(\frac{2d}{c+2d}\right)^{|C|-1}\cr
}
\tag{A.12}
$$
We will use these notations at many different places. 
As an example, let us prove formula (3.11).
Indeed, we have
$$
\eqalign{
&\left(\Pi_{\del G}\left(c-\D_{\L}\right)^{-1}\Pi_{\del G}\right)^{-1}
=c-\left(\D_{\del G}
+\del_{\del{G},\L\ba\del{G}}
R_{\L\ba\del{G}}
\del_{\L\ba\del{G},\del{G}}\right)\cr
&= c-\left(\D_{\del G}
+\del_{\del{G},\L\ba\del{G}}
R_{G^r\ba\del{G}}
\del_{\L\ba\del{G},\del{G}}\right)  
-\del_{\del{G},\L\ba\del{G}}
\left(R_{\L\ba\del{G}}-R_{G^r\ba\del{G}}\right)
\del_{\L\ba\del{G},\del{G}}\cr 
&=\left(\Pi_{\del G}\left(c-\D_{G^r}\right)^{-1}\Pi_{\del G}\right)^{-1}
-\del_{\del{G},\L\ba\del{G}}
\left(R_{\L\ba\del{G}}-R_{G^r\ba\del{G}}\right)
\del_{\L\ba\del{G},\del{G}}\cr 
&=\left(\Pi_{\del G}\left(c-\D_{G^r}\right)^{-1}\Pi_{\del G}\right)^{-1}
-\sum_{{C \sb \L\ba\del G}\atop
{C\cap(G^r)^c\neq \em,{C\cap G^2}\neq \em }}
\del_{\del{G},\L\ba\del{G}}\RR\left(\cdot\,\rightarrow\cdot\,; C\right)
\del_{\L\ba\del{G},\del{G}}\cr 
}
\tag{A.13}
$$
Here we have used Lemma A.1(i) in the first and third
equality and Lemma A.2 in the last one.
Finally we give the 

\proofof{Lemma 3.4}
The random walk representation of the determinant
is obtained writing
$$
\eqalign{
&\log \det\left(c-\D_{V}\right)= 
\log \det\left[\left( c + 2d\right)\left(
1+\frac{1}{c + 2d}T_{V}
\right)\right]\cr
&= |V|\log\left( c + 2d\right)
+\hbox{\rm Tr}\log \left(
1+\frac{1}{c + 2d}T_{V}
\right)\cr
}
\tag{A.14}
$$
and expanding the logarithm. 
Using (A.3) we can then write
$$
\eqalign{
&\log\frac{\det \left(\Pi_{\del G}\left(a-q\D_{G^r} \right)^{-1}
\Pi_{\del G}\right)}
{\det \left(\Pi_{\del G}\left(a-q\D_{\L} \right)^{-1}
\Pi_{\del G}\right)}
=\log \frac{\det_{G^r-\del G}\left(c-\D_{G^r-\del G}\right)}
{\det_{G^r}\left(c-\D_{G^r}\right)}
\frac{\det_{\L}\left(c-\D_{\L}\right)}
{\det_{\L-\del G}\left(c-\D_{\L-\del G}\right)}
\cr
&=\sum_{t=1}^\infty
\frac{1}{t}\frac{1}{(c+2d)^t}
\left(
\hbox{Tr}_{G^r}\left(T_{G^r} \right)^t
- \hbox{Tr}_{G^r-\del G}\left(T_{G^r-\del G} \right)^t
-\hbox{Tr}_{\L}\left(T_{\L} \right)^t
+ \hbox{Tr}_{\L-\del G}\left(T_{\L-\del G} \right)^t
\right)
\cr
}
\tag{A.15}
$$
It is not difficult to convince oneself that we have that 
$$
\eqalign{
&
\hbox{Tr}_{G^r}\left(T_{G^r} \right)^t
- \hbox{Tr}_{G^r-\del G}\left(T_{G^r-\del G} \right)^t
-\hbox{Tr}_{\L}\left(T_{\L} \right)^t
+ \hbox{Tr}_{\L-\del G}\left(T_{\L-\del G} \right)^t
\cr
&=-\sum_{x\in \L}\#\{
\g: x\mapsto x; \Range(\g)\sb \L;\Range(\g)\cap \del G\neq\em
; \Range(\g)\cap \L\ba G^r\neq\em;  |\g|=t \}\cr
}
\tag{A.16}
$$
So we get the form (3.28) putting
$$
\eqalign{
2\e^{\hbox{\srm det}}(C)
:=\sum_{t=2}^\infty
\frac{1}{t}\frac{1}{(c+2d)^t}
\sum_{x\in C}\#\{
\g: x\mapsto x; \Range(\g)=C; |\g|=t \}
}
\tag{A.17}
$$
From this the   
bounds of the form $\e^{\hbox{\srm det}}(C)\leq e^{-\const\left(\log c\right) 
|C|}$
are clear, assuming that $c$ is large.\endproof

\ftn
\font\bf=cmbx8

\baselineskip=11pt
\parskip=4pt
\rightskip=0.5truecm
\bigskip\bigskip
\chap{References}
\medskip


\item{[AW]} M.Aizenman, J.Wehr, Rounding Effects of Quenched Randomness
on First-Order Phase Transitions, Comm. Math.Phys {\bf  130},
489-528 (1990)

\item{[Ba1]} Balaban, Tadeusz,
A low temperature expansion for classical N-vector models. 
I. A renormalization group
flow, Comm.Math.Phys. {\bf 167} no. 1, 103--154 (1995)

\item{[Ba2]} Balaban, Tadeusz,
A low temperature expansion for classical N-vector models. 
II. Renormalization group
equations, Comm.Math.Phys. {\bf 182} no. 3, 675--721 (1996)




\item{[BCJ]} C.Borgs, J.T.Chayes, J.Fr\"olich,
Dobrushin States for Classical Spin Systems with 
Complex Interactions,
J.Stat.Phys. {\bf 89} no.5/6, 895-928 (1997)

\item{[BK1]} J.Bricmont, A.Kupiainen, 
Phase transition in the 3d random field Ising model,
Comm.
Math.Phys. {\bf 142}, 539-572 (1988)

\item{[BK2]} J.Bricmont, A.Kupiainen, 
High-Temperature Expansions and Dynamical
systems, Comm. Math.Phys. {\bf 178}, 703-732 (1996)

\item{[BKL]} J.Bricmont, A.Kupiainen, R. Lefevere,
       Renormalization Group Pathologies and the 
Definition of Gibbs States
Comm. Math.Phys. {\bf 194} 2, 359-388 (1998)
    
\item{[BoK]} A.Bovier, C.K\"ulske, A rigorous
renormalization group method for interfaces in random
media, Rev.Math.Phys. {\bf 6}, no.3, 413-496 (1994)

\item{[B]} D.Brydges, A short course on cluster expansions, 
in `Critical phenomena, random systems, gauge theories' 
(Les Houches 1984) (K.Osterwalder, R.Stora, eds.),
North Holland, Amsterdam (1986)



\item{[Do1]} R.L.Dobrushin, 
Gibbs states describing a coexistence of phases 
for the three-dimensional Ising model,
Th.Prob. and its Appl. {\bf 17}, 582-600 (1972)

\item{[Do2]} R.L.Dobrushin, 
Lecture given at the workshop `Probability 
and Physics', Renkum, August 1995

\item{[DoZ]} R.L.Dobrushin, M.Zahradnik, Phase Diagrams of Continuum Spin 
Systems, Math.Problems of Stat.Phys. and Dynamics, ed.
R.L.Dobrushin, Reidel pp.1-123 (1986)




\item{[EFS]} A.C.D.van Enter, R. Fern\'andez, A.Sokal,
Regularity properties and pathologies of position-space
renormalization-group transformations: Scope
and limitations of Gibbsian theory. J.Stat.Phys.
{\bf 72}, 879-1167 (1993)

\item{[FFS]} R.Fernandez, J.Fr\"ohlich, A.Sokal,
Random Walks, Critical Phenomena, and Triviality
in Quantum Field Theory, 
Springer, Berlin, Heidelberg, New York (1992)

\item{[Geo]} H.O. Georgii, Gibbs measures and phase transitions, Studies
in mathematics, vol. 9 (de Gruyter, Berlin, New York, 1988)

\item{[GJ]} J.Glimm, A.Jaffe, Quantum Physics:
A functional Integral Point of View, Springer, 
Berlin-Heidelberg-New York
(1981)

\item{[K]} C.K\"ulske, Ph.D. Thesis, Ruhr-Universit\"at Bochum (1994)


\item{[KP]} R.Kotecky, D.Preiss, Cluster expansion for
abstract polymer models, Comm.Math.Phys. {\bf 103},
491-498 (1986)


\item{[Na]} T.Nattermann, Theory of the Random Field Ising Model,
in `Spin Glasses and Random Fields', ed. P. Young, World
Scientific, available as cond-mat preprint 9705295 at 
http://www.sissa.it (1997)






\item{[S]} R.H.Schonmann, Projection of Gibbs measures may
be non-Gibbsian, Comm.Math.Phys. {\bf 124}
1-7 (1989)


\item{[Z1]} M.Zahradn\' \i k, `An alternate version
of Pirogov-Sinai theory', Comm.Math.Phys. {\bf 93}
559-581 (1984)

\item{[Z2]} M.Zahradn\' \i k, On the structure of low temperature
phases in three dimensional spin models with random impurities:
A general Pirogov-Sinai approach, Mathematical Physics
of Disordered Systems, Proc. of a workshop held at the CIRM,
ed. A.Bovier and F.Koukiou, IAAS-Report No.4, Berlin (1992)

\item{[Z3]} M.Zahradn\' \i k, Contour Methods and Pirogov
Sinai Theory for Continous Spin Models,
preprint Prague (1998),
to appear in the AMS volume dedicated to R.L.Dobrushin

\end